\documentclass[dvipsnames]{osa-article}
\usepackage{lipsum}
\usepackage{siunitx}
\usepackage{wrapfig}
\usepackage{booktabs}
\usepackage{placeins}
\usepackage{makecell}
\usepackage[textsize=footnotesize]{todonotes}
\usepackage{appendix}

\graphicspath{{figures/}}

\journal{osajournal}

\articletype{Research Article}

\begin{document}

\title{A Survey of Spatio-Temporal Couplings throughout High-Power Ultrashort Lasers}

\author{Antoine Jeandet\authormark{1,2,9}, Spencer W. Jolly\authormark{1,10}, Antonin Borot\authormark{1,11}, Beno\^it Bussi\`ere\authormark{2}, Paul Dumont\authormark{3}, Julien Gautier\authormark{4}, Olivier Gobert\authormark{1}, Jean-Philippe Goddet\authormark{4}, Anthony Gonsalves\authormark{5}, Wim  P. Leemans\authormark{5,12}, Rodrigo Lopez-Martens\authormark{4}, Gabriel Mennerat\authormark{1}, Kei Nakamura\authormark{5}, Marie Ouill\'e\authormark{4}, Gustave Pariente\authormark{1}, Moana Pittman\authormark{6}, Thomas Püschel\authormark{7}, Fabrice Sanson\authormark{8,2,13},  François Sylla\authormark{3}, C\'edric Thaury\authormark{4}, Karl Zeil\authormark{7}, and Fabien Qu\'er\'e\authormark{1,11,*}}

\address{\authormark{1}LIDYL, CEA, CNRS, Université Paris-Saclay, CEA Saclay, 91191, Gif-sur-Yvette, France\\
\authormark{2}Amplitude Laser Group, Business Unit Science, 2/4 rue du Bois Chaland, F-91090 Lisses, France\\
\authormark{3}SourceLAB, 7 rue de la Croix Martre, 91120 Palaiseau, France\\
\authormark{4}Laboratoire d’Optique Appliquée, CNRS, Ecole Polytechnique, ENSTA Paris, Institut Polytechnique de Paris, 181 chemin de la Hunière et des Joncherettes, 91120, Palaiseau, France\\
\authormark{5}Lawrence Berkeley National Laboratory, Berkeley, CA 94720, USA\\
\authormark{6}Laboratoire Ir\`ene Joliot-Curie, Universit\'e Paris-Saclay, CNRS, Rue Amp\`ere, B\^atiment 200, F-91898, Orsay Cedex, France\\
\authormark{7}Institute of Radiation Physics, Helmholtz-Zentrum Dresden-Rossendorf, 01328 Dresden, Germany\\
\authormark{8}LPGP, CNRS, Université Paris-Saclay, Bâtiment 210, 91400 Orsay, France\\
\authormark{9}Currently at Thales LAS France, 2 avenue Gay-Lussac, 78990 Elancourt, France\\
\authormark{10}Currently at Brussels Photonics (B-PHOT), Dept. of Applied Physics and Photonics, Vrije Universiteit Brussel, Pleinlaan 2, 1050 Brussels, Belgium\\
\authormark{11}Currently at Unistellar, 19 rue Vacon, 13001 Marseille, France\\
\authormark{12}Currently at Deutsches Elektronen-Synchrotron DESY, Notkestraße 85, 22607 Hamburg, Germany\\
\authormark{13}Currently at Imagine Optic, 18 Rue Charles de Gaulle, F-91400 Orsay\\
}

\email{\authormark{*}fabien.quere@cea.fr} %% email address is required

%% [use \begin{abstract*}...\end{abstract*} if exempt from copyright]
\begin{abstract}
The investigation of spatio-temporal couplings~(STCs) of broadband light beams is becoming a key topic for the optimization as well as applications of ultrashort laser systems. This calls for accurate measurements of STCs. Yet, it is only recently that such complete spatio-temporal or spatio-spectral characterization has become possible, and it has so far mostly been implemented at the output of the laser systems, where experiments take place. In this survey, we present for the first time STC measurements at different stages of a collection of high-power ultrashort laser systems, all based on the chirped-pulse amplification~(CPA) technique, but with very different output characteristics. This measurement campaign reveals spatio-temporal effects with various sources, and motivates the expanded use of STC characterization throughout CPA laser chains, as well as in a wider range of types of ultrafast laser systems. In this way knowledge will be gained not only about potential defects, but also about the fundamental dynamics and operating regimes of advanced ultrashort laser systems.
\end{abstract}

\section{Introduction}
Ultrashort laser sources are nowadays used in many different fields ranging from basic research to large-scale industrial applications. They can arguably be considered as some of the most advanced laser sources ever developed, as their proper operation and use call for very special care compared to other types of sources. Indeed, for the vast majority of applications, both their spatial and temporal degrees of freedom need to be accurately controlled and tailored: at focus, where these beams are generally used, the concentration of light energy is determined by the interference of many transverse (spatial) and longitudinal (frequency) modes, and thus ultimately by the relative phases of all these modes. Optimizing this energy concentration implies knowing and controlling this relative phase in a 3D space---two transverse positions, and time or frequency.
Very advanced methods have been available for years to accurately measure both the amplitude and phase of these beams in the spatial domain (transverse modes) on the one hand, and in the time/frequency domain (longitudinal modes) on the other hand. These measurement capabilities have been absolutely instrumental in the development and optimization of ultrashort laser sources. Yet, this metrology approach that treats spatial and temporal properties independently is inappropriate to characterize ultrashort pulses with arbitrary structures, such that the spatial and temporal properties are correlated---i.e., beams with so-called spatio-temporal couplings~(STCs)~\cite{akturk10}. This is a serious limitation, especially since Chirped Pulse Amplification~(CPA), the enabling technology of high-power ultrashort lasers, intrinsically makes use of massive STC, induced when broadband beams are diffracted by gratings or refracted by prisms, to be stretched or compressed in time. A complete understanding and optimization of high-power ultrashort lasers thus call for complete characterization in the space-time or space-frequency domains, ideally at different key stages of laser systems.

\begin{table}[h!]
\centering
\small
\caption{Main characteristics of the CPA lasers used for the measurement campaign, sorted by increasing peak power. All lasers have their spectrum centered around a wavelength of \SI{800}{\nano\meter}.}
{\setcellgapes{0.7ex}\makegapedcells
\begin{tabular}{@{}ccccccccc@{}}
\toprule
& Characteristics & \makecell{Measurement\\technique} & \makecell{Location of\\ measurements} & \emph{Main} application\\
\midrule
\makecell{\textbf{ARCO M}~\cite{ARCO}\\(Amplitude, FR)} & 0.1 TW, 20 fs, 1 kHz & TERMITES & \makecell{After stretcher,\\after $1^{st}$ amplifier} & \makecell{HHG,\\spectroscopy}\\
\makecell{\textbf{Salle Noire}~\cite{ouille20}\\(LOA, FR)} & 1 TW, <4 fs, 1 kHz & \makecell{TERMITES\\ \& INSIGHT} & \makecell{Before and after\\ post-compression}& \makecell{Few-cycle\\laser-plasma\\interactions}\\
\makecell{\textbf{UHI--100}\\(LIDYL, FR)} & 100 TW, 20 fs, 10 Hz & \makecell{TERMITES\\ \& INSIGHT} & Laser output& \makecell{Particle acceleration,\\harmonic generation \\from plasmas}\\
\makecell{\textbf{LASERIX}~\cite{Guilbaud:15}\\(LAL, FR)} & 40 TW, 40 fs, 10 Hz & TERMITES & Laser output & \makecell{Generation of\\X-ray lasers}\\
\makecell{\textbf{Salle Jaune}\\(LOA, FR)} & 150 TW, 30 fs, 10 Hz & \makecell{TERMITES\\ \& INSIGHT} & Laser output& \makecell{Laser wakefield\\ acceleration}\\
\makecell{\textbf{DRACO}~\cite{Schramm_2017}\\(HZDR, DE)} & \makecell{150 TW, 30 fs, 1 Hz,\\ and \\ 1 PW, 30 fs, 1 Hz\\} & INSIGHT & Laser outputs & \makecell{Particles acceleration}\\%,\\harmonic generation \\from plasmas}\\
\makecell{\textbf{BELLA}~\cite{nakamura17}\\(LBNL, USA)} & 1 PW, 40 fs, 1 Hz & \makecell{TERMITES\\ \& INSIGHT} & Laser output& \makecell{Laser wakefield\\acceleration}\\
\bottomrule
\end{tabular}}
\end{table} 

Such a complete insight into these laser systems has remained out-of-reach so far, due to the lack of suitable measurement techniques. Research work on STC metrology already started  about 20 years ago. But it is only recently that tractable techniques providing \textit{full} amplitude and phase information on the laser field \textit{in 3D} have become available and in common use~\cite{gabolde06,miranda14,parienteSpaceTimeCharacterization2016,borot18,dorrer19,jolly20-3}. Complete spatio-temporal characterization of ultrashort beams is thus about to be a routine metrology procedure, especially with commercial instruments now available \cite{Sourcelab}. Thanks to these recent developements, it is now becoming possible to fully examine high-power ultrafast lasers, not only at the system output but also from the inside, to achieve unprecedented understanding and optimization.

In this article, we present the first systematic experimental investigation of the spatio-temporal---or equivalently spatio-spectral---properties of ultrashort laser beams, at different stages of a variety of high-power laser systems. Our study reveals different types of complex effects that can be observed within these lasers, and that can affect their final performance. These clearly do not represent a comprehensive list of the spatio-temporal effects that can occur in all types of ultrashort lasers, and conversely we do not claim that these effects systematically occur in all such lasers. Yet, this provides for the first time a glimpse of the full spatio-temporal properties of high-power ultrashort laser pulses while they are being produced, and illustrates the potential of applying the latest generation of ultrafast diagnostics not only at the output of but also inside ultrashort laser systems. We hope that this will help open new trends in the development, control and optimization of ultrashort light sources---especially the most sophisticated ones, such as Optical Parametric Chirped-pulse Amplification~\cite{bromage2019}, Frequency-domain Optical Parametric Amplification systems~\cite{schmidt14}, systems with coherent combination~\cite{fsaifes20,stark21}, compressors based on tiled gratings~\cite{kessler04,li15,liu20}, or even more exotic scenarios to go beyond the Petawatt power range in future generations~\cite{li21}.

\begin{figure}[!ht]
\centering
\includegraphics[width=\linewidth]{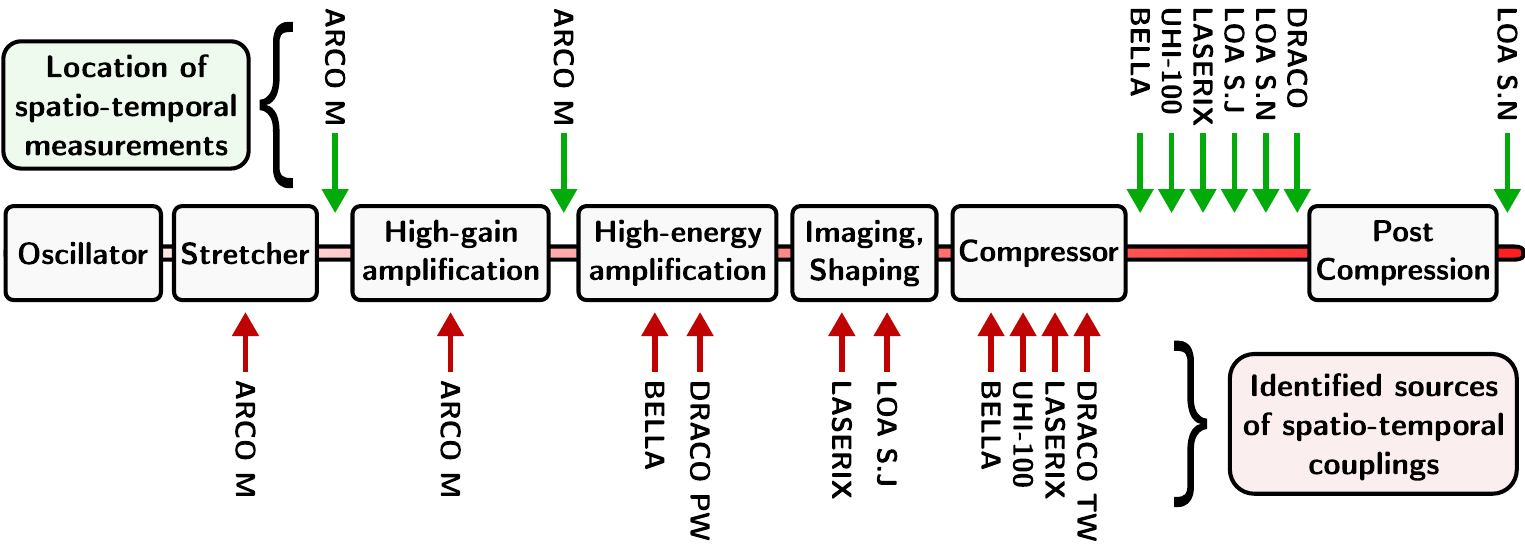}
\caption{\textbf{Summary of our STC survey.}
All lasers measured in this study (see Table~1) are based on the CPA technique, and hence share the same overall architecture, summarized by the central line of this sketch. The upper line indicates where we performed STC measurements on these different systems. The lower line shows the sources of STC that we have been able to identify through these measurements.}
\label{fig:survey-summary}
\end{figure}

To carry out this investigation, we used two spatio-temporal measurement methods developed in the last 5 years, TERMITES and INSIGHT. The principle and main features of these methods are summarized in appendix \ref{sec:measurement-techniques}. The key point here is that these two techniques can provide the complex (i.e., amplitude and phase) 3D spatio-temporal field, $E(x,y,t)$, or spatio-spectral field, $\Tilde{E}(x,y,\omega)$, in any arbitrary plane along the propagation axis. In this article, we will always display this information in the near-field, corresponding to a collimated beam far from a focus, and \emph{typically} in the frequency domain rather than in time, because this is generally the representation that lends itself most easily to interpretation. Representations in space-time will also be provided when this proves useful.

These methods, which do not require temporally-compressed beams, were implemented at different stages of six high-power laser systems, operated in different laboratories/companies worldwide and listed in Table~1. Although these lasers have very different output characteristics, they all use  Titanium-Sapphire as the amplifying medium, and are all based on the CPA technology, and hence have the same overall architecture, sketched in Fig.~\ref{fig:survey-summary}: the beam from a broadband oscillator is stretched in time, before being amplified in a series of amplifiers and transversally expanded, and finally compressed temporally. Note that the way this technology is implemented can actually strongly differ from one system to the other. We will present, discuss and analyze the results of spatio-temporal measurements carried out at a few locations and linked to all stages of these systems (see Fig.~\ref{fig:survey-summary}). Although still most of the measurements were done at the final ouput of each system, we reveal spatio-temporal effects coming into play from the output of the pulse stretcher (section \ref{sec:spatio-temporal-effects-of-pulse-stretching}), to the output of the different amplifiers (section \ref{sec:spatio-temporal-effects-of-pulse-amplification}), along relay imaging lines inside or after the laser system (section \ref{sec:spatio-temporal-effects-of-beam-imaging}), after the compressor (i.e., laser output, section \ref{sec:spatio-temporal-effects-of-pulse-compression}), and after a hollow-core fiber post-compression stage (section \ref{sec:spatio-temporal-effects-of-hollow-core-fiber-compression}).

\section{Effects of Pulse Stretching}
\label{sec:spatio-temporal-effects-of-pulse-stretching}

The pulse stretcher is a key component of any CPA laser chain. It is used to stretch femtosecond pulses in time up to the 100's of picoseconds scale, thus reducing the laser peak power by orders of magnitude before the subsequent amplification steps. This is ideally achieved without affecting the spectral content of the pulse, by adding strong amounts of spectral dispersion~\cite{cheriauxAberrationfreeStretcherDesign1996}. Because multiple components of pulse stretchers induce massive amounts of STC (namely angular dispersion and spatial chirp), these systems have to be very carefully designed and aligned, such that these multiple sources of STC compensate each other as accurately as possible at the output and the net effect of this device is to introduce a phase which is purely spectral, free of any spatio-spectral couplings.

\begin{figure}[htbp]
\centering
\includegraphics[width=\linewidth]{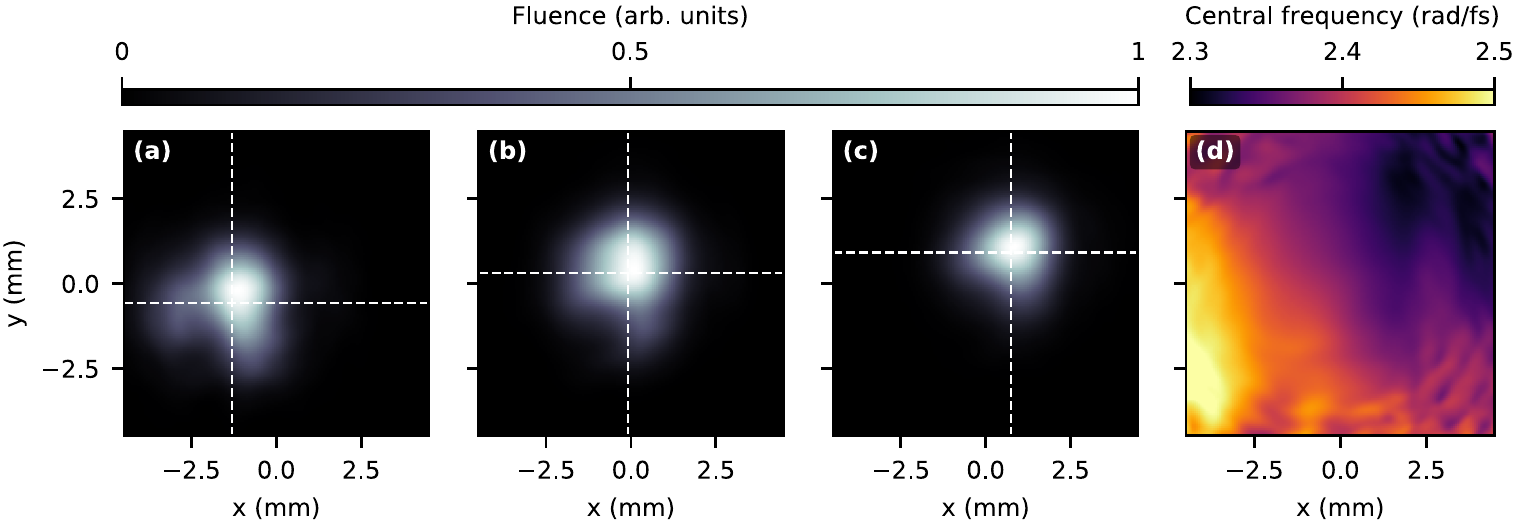}
\caption{\textbf{Spatial Chirp measured at the output of an Offner stretcher.} \mbox{\textbf{(a)--(c)}}~Spectrally-resolved spatial profiles of the beam at $2.5$, $2.375$ and \SI{2.25}{\radian/\femto\second} (respectively corresponding to wavelengths of $754$, $793$ and \SI{837}{\nano\meter}). Centroids are computed for each profile, and represented as white crosses. \textbf{(d)}~Color map of the central angular frequency of the pulse as a function of the transverse position in the beam. This is determined by computing the barycenter of the spectrum at each transverse spatial position. The angular dispersion of the stretcher's gratings is along the x-axis.}
\label{fig:spatial-chirp-stretcher}
\end{figure}

In this first section, we illustrate how spatio-spectral diagnostics can provide information on the potential misalignements of stretchers and their resulting effects on the laser beam when sent to amplifiers. This measurement was performed with TERMITES on a medium size laser chain under construction in Amplitude Laser laboratories (first line in Table~1), which uses a standard single-grating \emph{Offner} stretcher. This stretcher was prepared using the standard alignment procedure used in most laboratories, which did not reveal any significant misalignement.

Fig.~\ref{fig:spatial-chirp-stretcher} displays the results of a measurement that was done immediately after this Offner stretcher. The output beam exhibits transverse spatial chirp: panels~\mbox{(a)--(c)} show that different wavelengths produce beams that are centered at different tranverse positions.
The spatial shape of the beam does not vary significantly with wavelength, indicating the absence of higher-order amplitude couplings. Furthermore, since the measurement also provides the beam wavefront at each frequency, we have been able to check that the beam has a negligible angular dispersion, i.e., the different frequencies, although sheared spatially, all propagate in the same direction at the stretcher's output.

\begin{figure}[htb]
\centering
\includegraphics[width=0.7\linewidth]{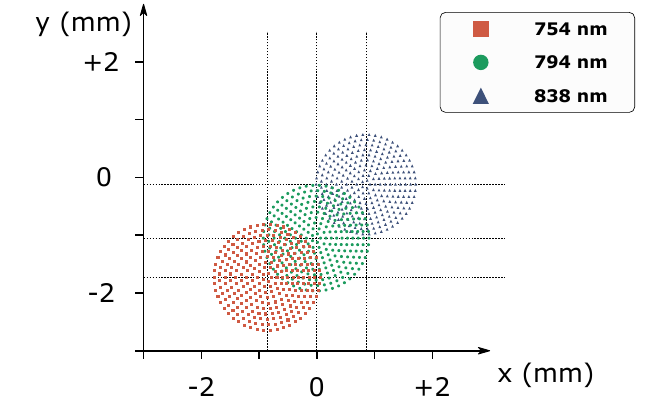}
\caption{\textbf{Simulated spot diagrams at the output of a misaligned Offner stretcher}
These spots diagrams were calculated with Zemax, at three different frequencies (blue: \SI{754}{\nano\meter}, green: \SI{794}{\nano\meter}, red: \SI{838}{\nano\meter}), considering a misaligned Offner stretcher with the parameters provided in Table~2. The results are quantitatively consistent with the experimental results displayed in Fig.~\ref{fig:spatial-chirp-stretcher}.}
\label{fig:zemax-stretcher}
\end{figure}

These spectrally-resolved spatial profiles provide a very intuitive way of visualizing the spatial dispersion of the pulse frequency components, which amounts here to \SI{10}{\milli\meter/(\radian/\femto\second)}. This is relatively strong, considering the diameter ($\simeq\SI{3}{\milli\meter}$) and the spectral bandwidth (\SI{100}{\nano\meter} or \SI{0.3}{(\radian/\femto\second)}) of the beam at this point of the laser system. Panel~(d) of Fig.~\ref{fig:spatial-chirp-stretcher} displays the same information in a different, more synthetic way: it shows, in color code, the spatial dependence of the central angular frequency of the measured beam. This clearly reveals that the measured spatial chirp is monotonous and oblique (i.e., not along the vertical or horizontal axis only).

\begin{table}[htb]
\centering
\small
\caption{Parameters of the Zemax simulation of a slightly-misaligned single-grating Offner stretcher, and main results}
{\setcellgapes{0.9ex}\makegapedcells
\begin{tabular}{@{}cc@{}}
\toprule
Component & Characteristics \\
\midrule
\makecell{\textbf{Grating}} & \makecell{1200 lines/mm \\Angle of incidence \SI{37}{\degree}}\\
\makecell{\textbf{Concave mirror}} & \makecell{Radius of curvature \SI{1}{\meter}} \\
\makecell{\textbf{Distance grating-concave mirror}} & \SI{0.75}{\meter}\\
\textbf{Convex mirror} & \makecell{Radius of curvature \SI{0.5}{\meter}\\Deviations from perfect alignment:\\Vertical shift by \SI{1}{\milli\meter}\\Rotation around vertical axis by \SI{0.25}{\degree}\\ Rotation around horizontal axis by \SI{-1}{\degree}}\\
\midrule
\makecell{\textbf{Main results}} & \makecell{Spatial chirp \SI{9.5}{\milli\meter/(\radian/\femto\second)} \\ Negligible angular dispersion ($<\SI{1}{\micro\radian/\nano\meter}$)}\\
\bottomrule
\end{tabular}}
\end{table}

In the case of single-grating Offner stretchers, as considered here, a small non-linear vertical spatial chirp on the output beam is expected even for a perfectly aligned system. However, the \textit{linear} spatial chirp observed here in the vertical direction is much stronger than this residual defect of an ideal system. This fact, as well as the observation of a spatial chirp in the horizontal direction, rather point to a misalignement of the stretcher. We used ray-tracing simulations of an Offner stretcher with Zemax to check this interpretation and quantify the involved misalignement. We were indeed able to reproduce spatial chirps of magnitudes comparable to the measured ones in both directions (Fig.~\ref{fig:zemax-stretcher}), by considering slight misalignements of the stretcher's convex mirror (see Table~2), well within the accuracy of standard alignement procedures.

In the next section, we will follow the evolution of the spatially-chirped beam produced by this slightly-misaligned stretcher as it passes through the first amplifier of the laser system, of regenerative type.

\section{Effects of Pulse Amplification}
\label{sec:spatio-temporal-effects-of-pulse-amplification}
This section is dedicated to the effects of the different amplification stages of high-power femtosecond lasers on the spatio-spectral properties of the beam. We will start by presenting the results obtained at the output of the first high-gain low-energy amplification stage of the laser system already considered in the previous section. We will then consider the very different case of a PW laser beam, where a significant amplitude spatio-spectral coupling is observed on the final beam, induced by the last high-energy low-gain multipass amplification stage used in this type of system. 

\subsection{Spatio-spectral Filtering by Amplification}
\label{subsec:spatio-spectral-filtering-by-amplification}

The characterization done in section~\ref{sec:spatio-temporal-effects-of-pulse-stretching} was immediately followed by measurements at the output of the regenerative amplifier, located directly after the Offner stretcher. Both sets of measurements were performed on the same day, in identical conditions, such that the beam injected in the amplifier was affected by the spatial chirp induced by the slightly-misaligned stretcher, described in section \ref{sec:spatio-temporal-effects-of-pulse-stretching}.

\begin{figure}[htb]
\centering
\includegraphics[width=\linewidth]{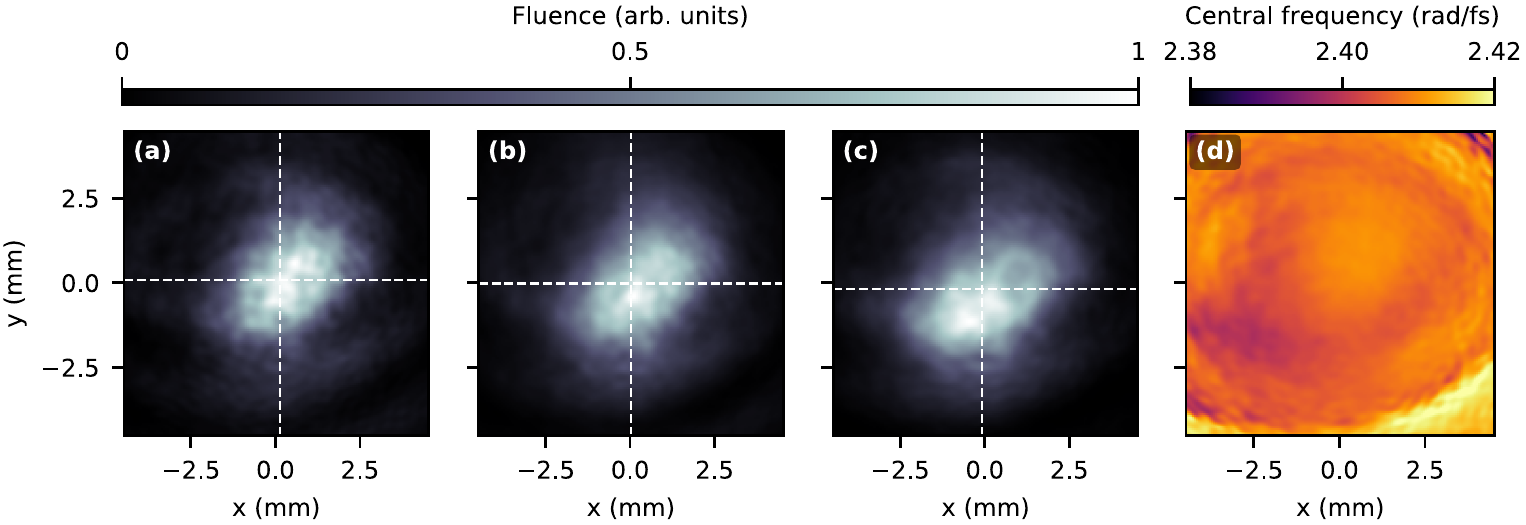}
\caption{\textbf{Spatial Chirp measured at the output of a regenerative amplifier seeded by a misaligned stretcher.} \mbox{\textbf{(a)--(c)}}~Spectrally-resolved spatial profiles of the beam at $2.47$, $2.41$ and~\SI{2.35}{\radian/\femto\second} (respectively corresponding to wavelengths of $763$, $782$ and \SI{802}{\nano\meter}). Centroids are computed for each profile and represented as white crosses. Because of the spectral narrowing occurring in the amplifier (see Fig.~\ref{fig:regen-spectra}(a)), the displayed frequencies are different from those in Fig.~\ref{fig:spatial-chirp-stretcher}. \textbf{(d)}~Color map of the central angular frequency of the pulse as a function of the transverse position in the beam. The spatial chirp is clearly weaker than at the stretcher output (note that the color scale is different from that of Fig.~\ref{fig:spatial-chirp-stretcher}(d)).
}
\label{fig:spatial-chirp-amplifier}
\end{figure}

\begin{figure}[htb]
\centering
\includegraphics[width=\linewidth]{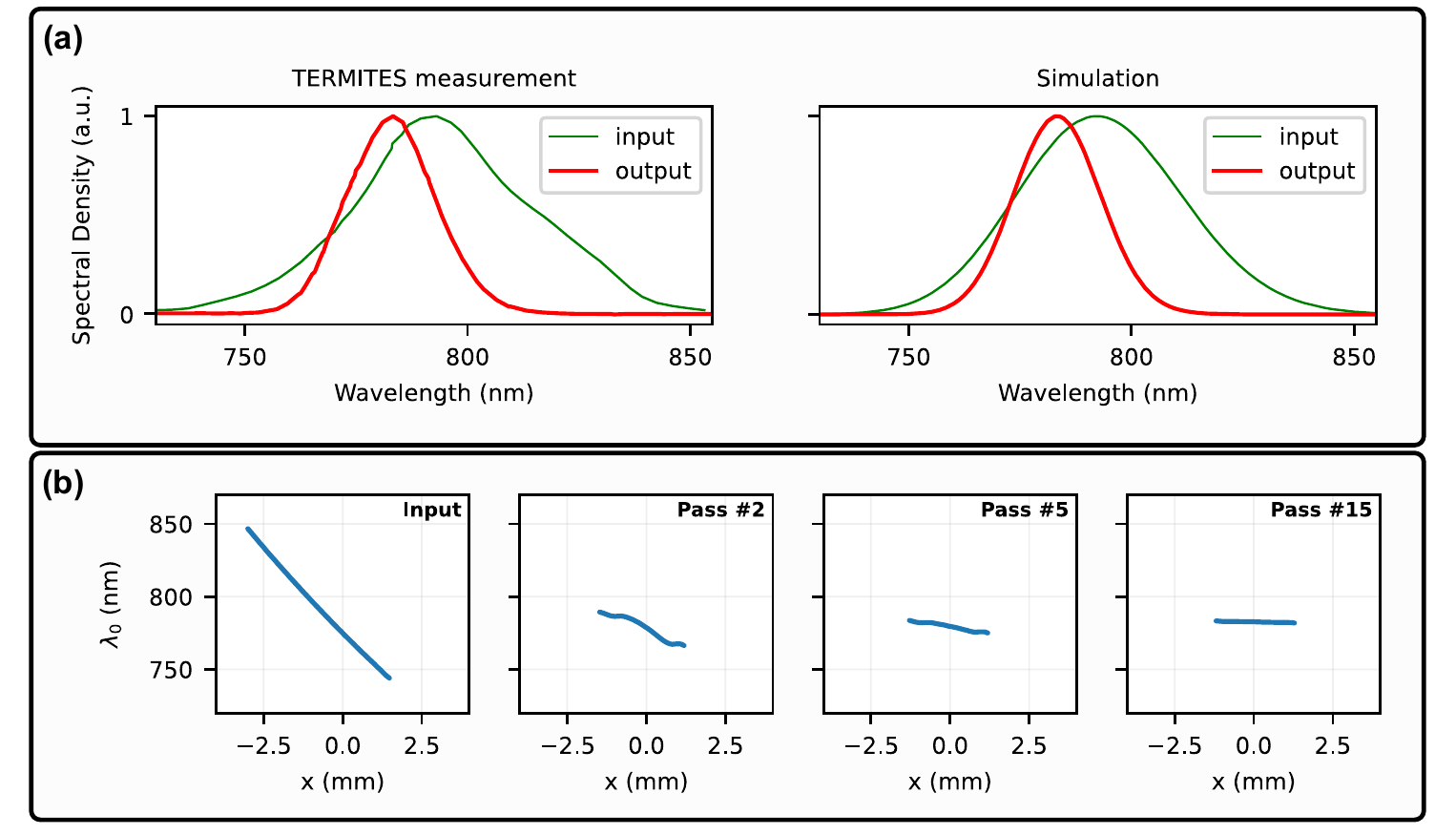}
\caption{\textbf{Output of a regenerative amplifier for a given input spatial chirp.} When spatial chirp and an additional constant spatial offset of the seed beam are included \textbf{(a)} both the spectral narrowing and slight blueshift of the output spectrum seen in measurements (left) can be reproduced via simulations (right). The central wavelength as a function of transverse position is shown after different number of passes \textbf{(b)} for an initial spatial chirp of \SI{6}{\milli\meter/(\radian/\femto\second)} from the simulations.}
\label{fig:regen-spectra}
\end{figure}

Fig.~\ref{fig:spatial-chirp-amplifier} displays the measurement data in the same manner as the one we used previously to show the spatial chirp generated by the Offner stretcher. Comparing this figure to Fig.~\ref{fig:spatial-chirp-stretcher} immediately shows that spatial chirp has been almost totally suppressed by the regenerative amplifier. Another striking effect is the significant spectral narrowing induced by the amplifier, shown in the left panel of Fig.~\ref{fig:regen-spectra}(a).
A spectral narrowing by amplifiers is always expected, and is generally attributed to the finite spectral width of the amplifying medium's gain---the so-called spectral gain narrowing. Using the support of numerical simulations, we are now going to show that spatio-spectral effects actually also play a very significant role in the observed spectral narrowing: more precisely, the initial spatial chirp is suppressed because the regenerative cavity acts as a non-linear spatial filter, but the price to pay for this filtering is an increased reduction of the pulse spectral width after amplification.

The numerical simulation framework used to describe the regenerative amplifier is sumarized in Appendix B. By injecting a beam in the cavity with a spatial chirp close to the one measured at the stretcher output, these simulations can satisfactorily reproduce the large spectral narrowing seen in the experiments (compare right and left graphs of Fig.~\ref{fig:regen-spectra}(a)). The slight blueshift of the central frequency observed experimentally is also reproduced. We note that  regenerative amplifiers are generally rather expected to cause a redshift in the spectra. But here the combination of an initial spatial chirp of \SI{6}{\milli\meter/(\radian/\femto\second)} and a slight transverse spatial offset of the beam with respect to the cavity axis (which we have adjusted to $\delta x=0.45$\,mm to fit the experimental results) cause preferential amplification of the blue part of the spectrum. 
Remarkably, we also observe in these simulations that the significant initial spatial chirp of the beam (first panel of Fig.~\ref{fig:regen-spectra}(b) is progressively reduced with each pass until there is effectively zero residual spatial chirp after 15 passes (last panel of Fig.~\ref{fig:regen-spectra}(b)). The two effects observed experimentally are thus reproduced, therefore we can now exploit the simulations to analyze their causes.

\begin{figure}[htb]
\centering
\includegraphics[width=\linewidth]{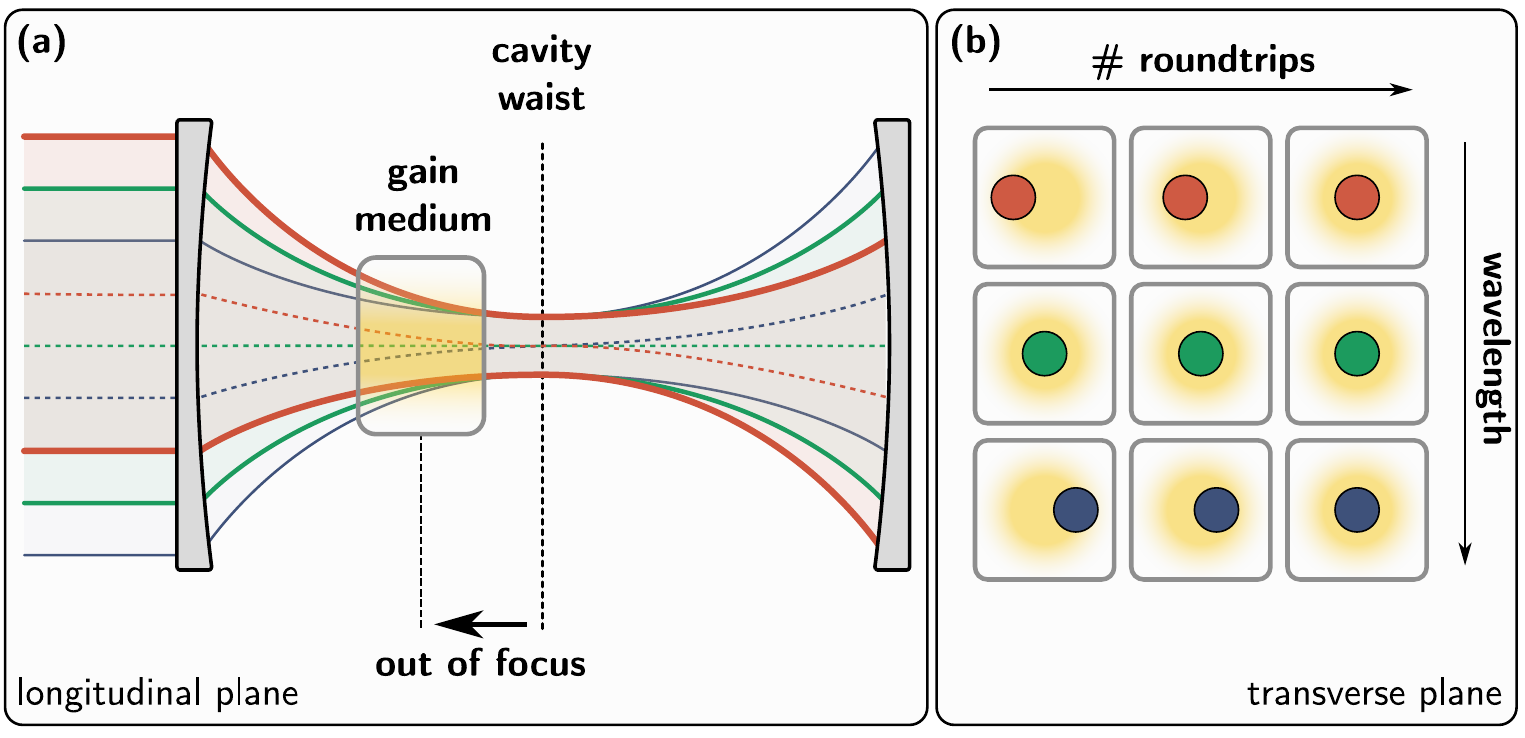}
\caption{\textbf{Amplification of a spatially chirped beam in a regenerative amplifier.} Panel~(a) features a simplified regenerative amplifier cavity, wherein Pockels cells and polarizers are omitted. Because gain medium is positioned slightly out-of-focus, some residual spatial chirp affects the amplified pulses when they propagate through the crystal. Panel~(b) is a simple sketch of the spatial chirp filtering occuring at each pass in this configuration. In both panels, locations colored in yellow represent the pumped section of the crystal.}
\label{fig:regen-sketch}
\end{figure}

The combination of the amplifier geometry, visualized in Fig.~\ref{fig:regen-sketch}(a), with a spatio-spectrally aberrated seed pulse has two interrelated consequences on the amplified pulse. The first consequence is purely spectral. In the presence of transverse spatial chirp in the beam, only a fraction of the seed spectrum  efficiently couples to the amplifying optical cavity, and this thus introduces a spectrally-dependant gain in the system. This effect results in the narrowing of the beam spectrum, and should not be confused with spectral gain narrowing, which is linked to the gain bandwidth of the amplifier crystal. 

The second effect, and the reason for the decreasing spatial-chirp at the output of the amplifier, shown in Fig.~\ref{fig:regen-spectra}(b) and measured in the experiment, is the fact that not only is the loss spectrally-dependent when the input is spatially-chirped, but the eventual mode that each wavelength settles on is related to the single transverse cavity mode. As visualized in Fig.~\ref{fig:regen-sketch}(b), the colors that are originally off-center (i.e. not overlapping with the cavity mode) will tend towards the cavity mode after a few passes. This effect clearly appears in movie S1 (see Supplementary Material) of the spatial profile of the beam over multiple passes, obtained from the simulations, where all frequencies are observed to gradually collapse to the same transverse spatial profile as the number of passes increases.

\begin{figure}[htb]
\centering
\includegraphics[width=\linewidth]{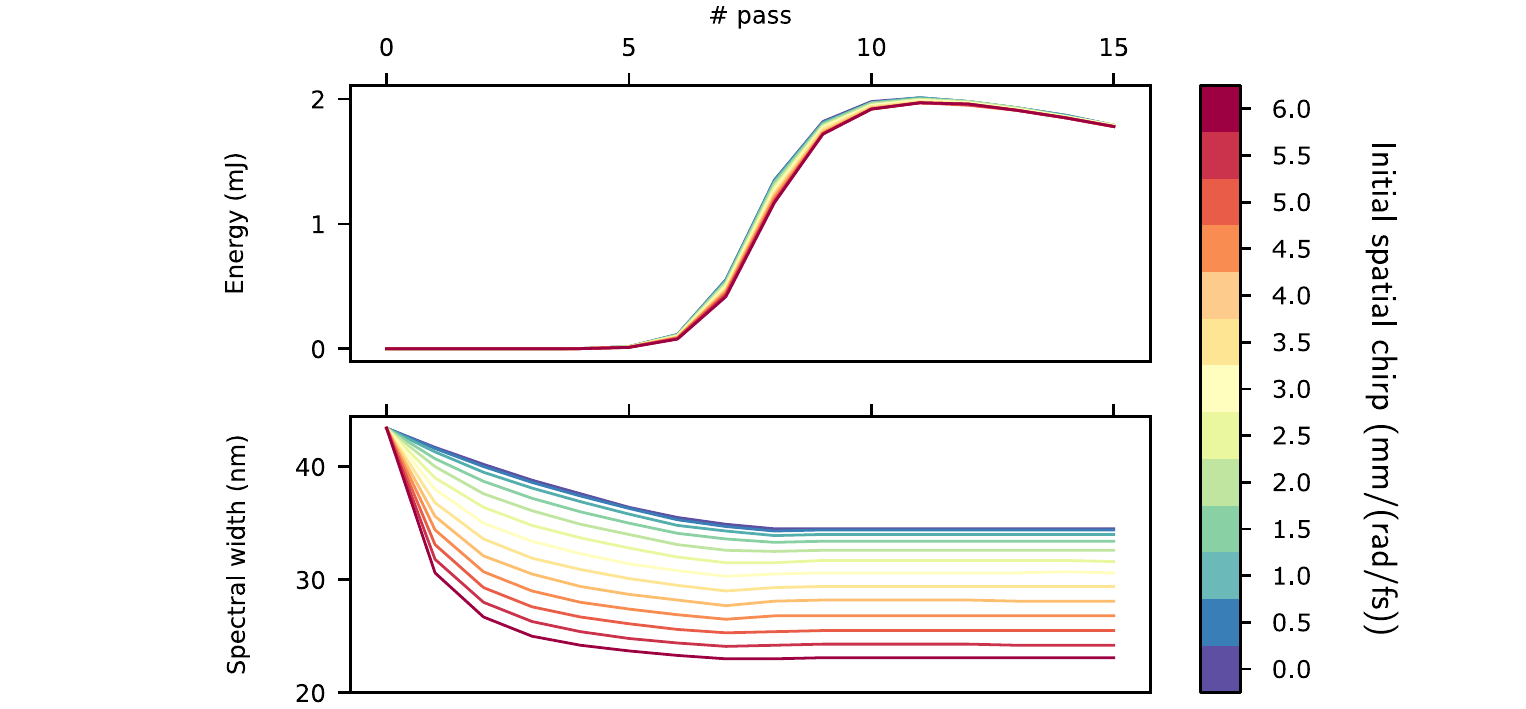}
\caption{\textbf{Performance of a regenerative amplifier with a spatially-chirped seed.} The pulse energy (top) and spectral bandwidth (bottom) are shown after each pass of a simulated regenerative amplifier for different values of the intial spatial chirp (colorbar on the right). While the final energy is relatively insensitive to the initial spatial chirp, the final bandwidth gets strongly reduced as the initial spatial chirp increases, beyond the reduction due to only gain-narrowing.}
\label{fig:regen-sim}
\end{figure}

To investigate more quantitatively the impact of an initial spatial chirp on the output of a regenerative amplifier, we have also performed simulations pass-by-pass for different values of initial spatial chirp, shown in Fig.~\ref{fig:regen-sim}. The first result in Fig.~\ref{fig:regen-sim}(top) is that even with different values of initial spatial chirp the output energy of the amplifier after each pass is relatively constant. This means that in experimental conditions where the output energy is the first figure of merit for optimizing the amplifier, these effects would not be seen, as the same energy is always extracted from the cavity whatever the initial spatial chirp. 

However, the spectral width decreases much more significantly as the initial spatial chirp is increased. This narrowing mostly increases with the first few passes, which is fully consistent with the interpretation of the previous paragraphs. Fig.~\ref{fig:regen-sim}(bottom) shows that with an initial spectral width of 43\,nm and no initial spatial chirp, the spectral width decreases to 35\,nm, entirely due to spectral gain narrowing. With an initial spatial chirp of \SI{6}{\milli\meter/(\radian/\femto\second)}, the spectral width reduces to 23\,nm, \SI{35}{\percent} smaller than the case with no initial spatial chirp. This demonstrates the very significant impact of initial transverse spatial chirp on spectral narrowing in regenerative amplifiers.

\FloatBarrier
\subsection{Gain Depletion in High-energy Chirped Pulse Amplifiers}
\label{subsec:gain-depletion-in-chirped-pulse-amplifiers}
\index{Chirped Pulse Amplification}

The next defect that will be studied in this survey is a spatio-spectral amplitude coupling induced during the amplification process, specifically related to the last, highest energy amplification stage of a Petawatt laser system~\cite{jeandetSpatiotemporalStructurePetawatt2019}.

\begin{figure}[htb]
\centering
\includegraphics[width=\linewidth]{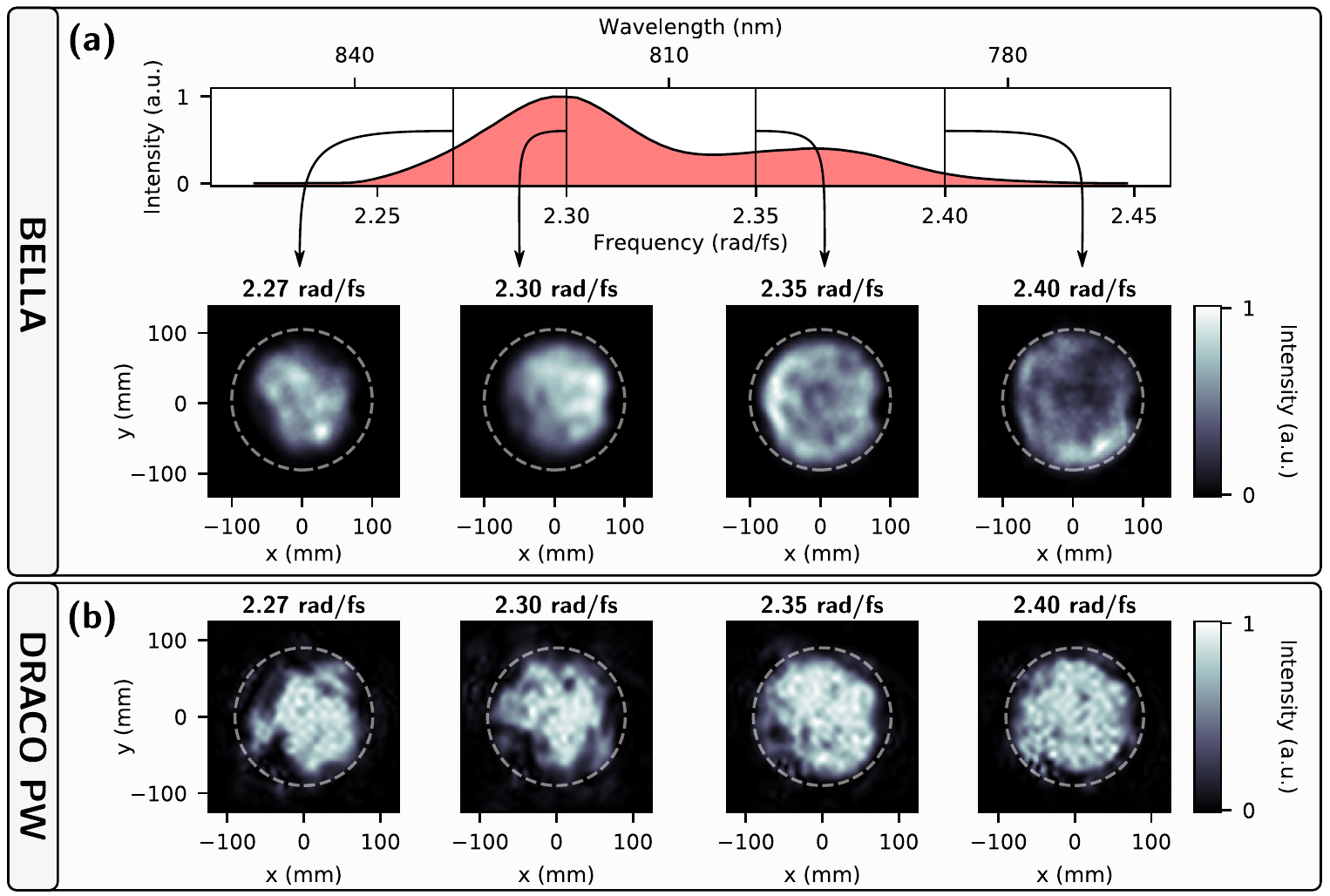}
\caption{\textbf{Spectrally-resolved spatial profile of the full power BELLA and DRACO PW beams.} Spectrally-resolved spatial intensity profiles, taken at respectively $830$, $820$, $800$ and \SI{785}{\nano\meter} from left to right, with the BELLA beam in (a) and the DRACO PW beam in (b). The upper panel in (a) shows the spatially-integrated spectrum of the measured BELLA pulse. The dashed white circle has a constant diameter for each laser system, and provides a scale reference. To clearly highlight the spectral evolution of the beam spatial profile, all spatial profiles have been normalized so that the energy is the same for each spectral sample. The data in (a) is from Ref.~\cite{jeandetSpatiotemporalStructurePetawatt2019} \copyright IOP Publishing.}
\label{fig:bella-spectral-fluence}
\end{figure}

Fig.~\ref{fig:bella-spectral-fluence}(a) illustrates the spectrally-resolved spatial intensity profile of the BELLA laser beam operating at full power. This picture shows that the beam profile is not spectrally constant, but evolves smoothly with laser frequency. In particular, it should be noted that the beam is top-hat around \SI{820}{\nano\meter} (red side of the spectrum), and evolves to a ring shape around \SI{780}{\nano\meter} (blue side of the spectrum). Furthermore, the diameter is also smaller for larger wavelengths before increasing for lower wavelengths. Fig.~\ref{fig:bella-spectral-fluence}(b) shows similar data at the same four frequencies for the DRACO PW system at full power, showing as well the increase in beam size with frequency.

As this effect could not be explained by well-known sources of spatio-spectral couplings---like a stretcher/compressor or lenses---we looked for the culprit elsewhere in the laser chain, by studying in details the amplification process taking place in the successive Ti:Sapphire amplifiers of the BELLA laser. This analysis relied on the Frantz-Nodvik equations~\cite{frantzTheoryPulsePropagation1963}, which model the energy transfer from the population inversion of the medium (pumped by frequency-doubled Nd:YLF and Nd:YAG) to the amplified beam. They were used to compute numerically the profiles of the gain medium and of the amplified beam at each passage through the amplifier.

\begin{figure}[htb]
\centering
\includegraphics{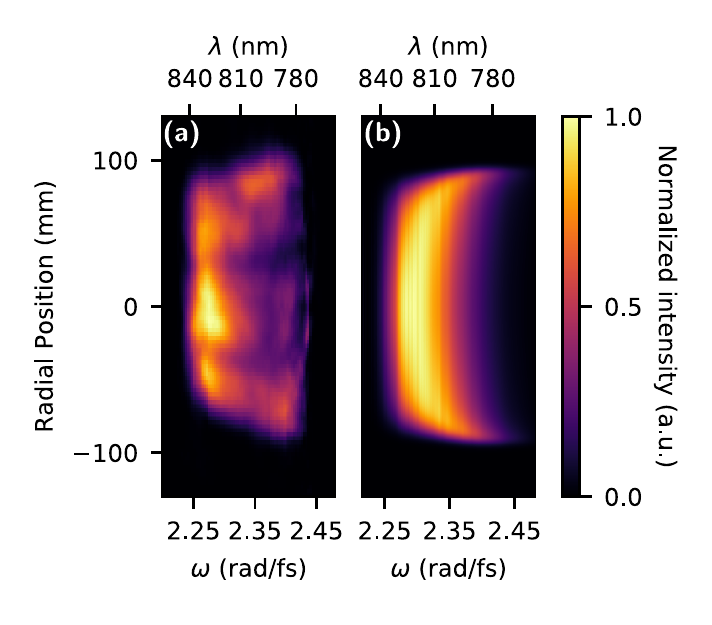}
\captionsetup{format=plain}
\caption{\textbf{Spatio-spectral intensity of the full power BELLA beam.} \textbf{(a)}~$(x, \omega)$ slice of the spatio-spectral intensity as measured by INSIGHT, taken across the horizontal plane and vertically centered. \textbf{(b)}~Same quantity, now deduced from simulations of the CPA process based on the Frantz--Nodvik equations~\cite{frantzTheoryPulsePropagation1963}.
%Both panel has been normalized following Eq.~\ref{eq:bella-profile-normalization}. 
This data is from Ref.~\cite{jeandetSpatiotemporalStructurePetawatt2019} \copyright IOP Publishing.
}
\label{fig:bella-mode-size}
\end{figure}

Fig.~\ref{fig:bella-mode-size} shows two $x-\omega$ slices of BELLA spectral fluence side-by-side. The one in panel~(a) was generated from the amplification simulations, while the one in panel~(b) is a slice of the quantity obtained experimentally with INSIGHT. This figure shows that the simulation is qualitatively very close to the experimental measurement. Both slices feature top-hat modes for low frequencies and ring shape for high frequencies, while the overall beam diameter evolves linearly from $170$ to \SI{200}{\milli\meter}. Note that no such amplitude coupling is predicted via simulations in earlier amplifiers (consistent with the spatio-spectral filtering effect described in the previous subsection), implying that it is probably only present in the highest energy CPA systems. Although in the case of DRACO PW, Fig.~\ref{fig:bella-spectral-fluence}(b),  there is no ring shape at the highest frequency, this depends on the exact state of the laser system.

\begin{figure}[htb]
\centering
\includegraphics[width=0.78\linewidth]{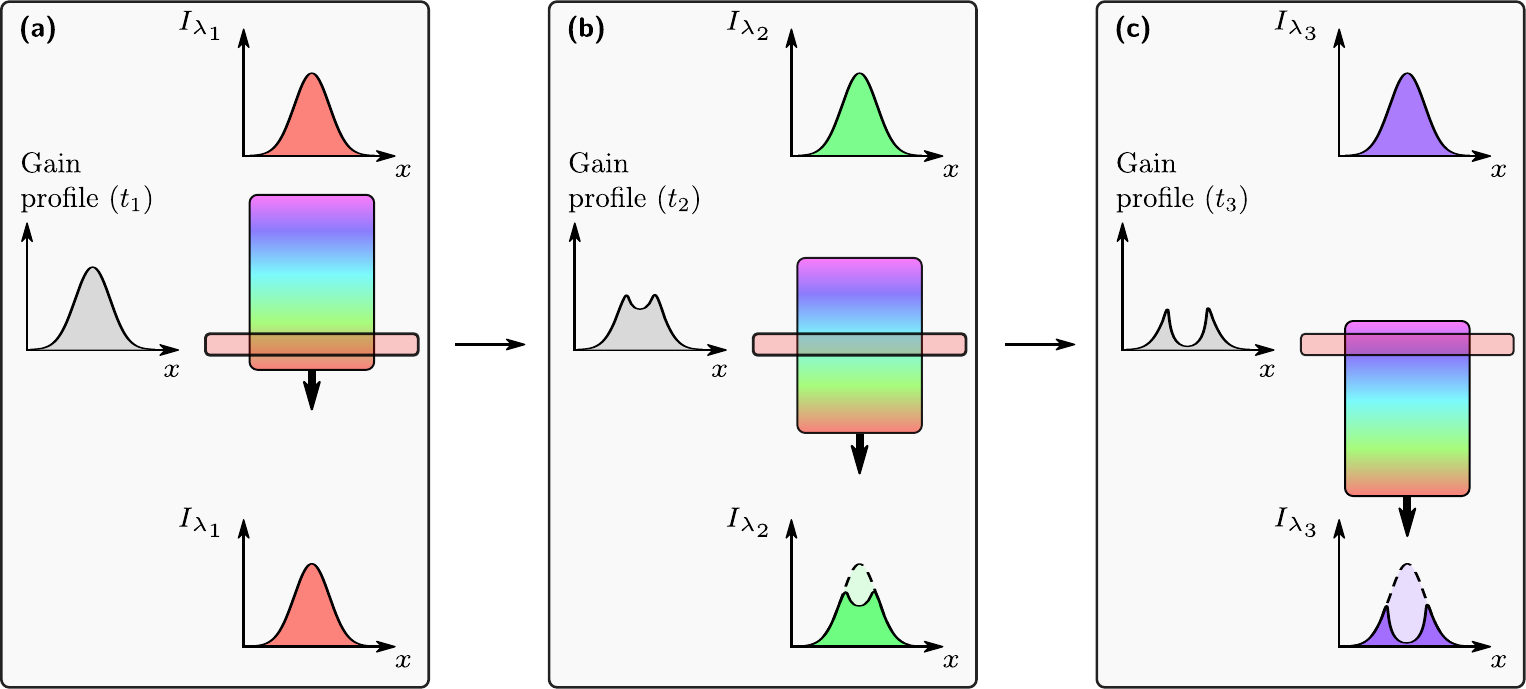}
\caption{\textbf{Sketch of a chirped pulse going through the gain medium of a laser amplifier operated in a saturated regime.} This drawing shows a chirped pulse in the process of being amplified by a pumped Ti:Sapphire crystal. \textbf{(a)}~Because the input beam is positively chirped, red components of the laser spectrum are amplified ahead. \mbox{\textbf{(b)--(c)}}~As shorter wavelengths go through the amplifying medium, the population inversion profile (transverse gain profile) is more depleted at the center than on the edges. Output fluence profiles at those shorter wavelengths are therefore not Gaussian anymore.
}
\label{fig:beam_size_explanation}
\end{figure}

This spatio-spectral amplitude coupling is a direct side-effect of the Chirped Pulse Amplification technique (Fig.~\ref{fig:beam_size_explanation}).
The beam profile after amplification is a function of the seed profile and of the population inversion (or stored energy) profile in the gain medium.
Considering that both quantities have radial Gaussian distributions at the beginning of the laser chain is a reasonable approximation that was used for the numerical simulations (panel~(a) in Fig.~\ref{fig:beam_size_explanation}). When a chirped pulse is amplified and the amplifier is in a saturated regime, its leading edge (red part of the spectrum) depletes a fraction of the energy stored in the amplification medium (panels~(a)--(b) in Fig.~\ref{fig:beam_size_explanation}). This depletion effect is stronger at the center of the beam than on its edges, due to the stronger seed fluence. This thus affects the spatial distribution of population inversion \emph{seen} by lower wavelengths arriving at later times (panels~\mbox{(b)--(c)}) and hence modifies the spatial profile of those components.  This spatio-spectral amplitude coupling is thus strongly linked to the CPA scheme: it would not happen if all frequencies were amplified at the same time. It can be considered as a spatio-spectral extension to the traditional CPA-specific red-shift effect~\cite{durfeeDesignImplementationTWclass1998,giambrunoDesign10PW2011}. Red-shift in CPA lasers is indeed well-understood, but has so far been described solely in the spectral domain to the best of our knowledge.

\section{Effects of Beam Imaging}
\label{sec:spatio-temporal-effects-of-beam-imaging}

In all high-power ultrashort lasers, the beam needs to be expanded in diameter as it gets amplified along the system. In some lasers, relay imaging systems are also used between sub-modules, to avoid the build-up of strong fluence modulations on the beam, which could lead to damage of some optical elements. All these operations are very often implemented using telescopes based on lenses. In this section, we present multiple measurements of STC typically induced by such systems, which most often consist of Pulse Front Curvature (PFC) induced by chromatic lenses. 

PFC is a spatio-temporal amplitude coupling which adds a delay between the phase front and the pulse front of the beam. This delay depends quadratically on the distance from the center of the beam, curving the pulse front spatio-temporally relative to the wavefront. PFC is the temporal consequence of chromatic curvature, whereby the wavefront curvature depends linearly on angular frequency. The most straightforward way to add PFC to an ultrashort pulse is to use uncompensated singlet lenses~\cite{borDistortionFemtosecondLaser1988,borDistortionFemtosecondLaser1989}. Because of the widespread use of large singlet lenses in beam expanding telescopes, and since chromatic curvature and PFC are equivalent in ultrashort pulses, beam imaging in broadband laser systems can significantly affect the spatio-temporal properties.

\begin{figure}[htb]
\centering
\includegraphics[width=\linewidth]{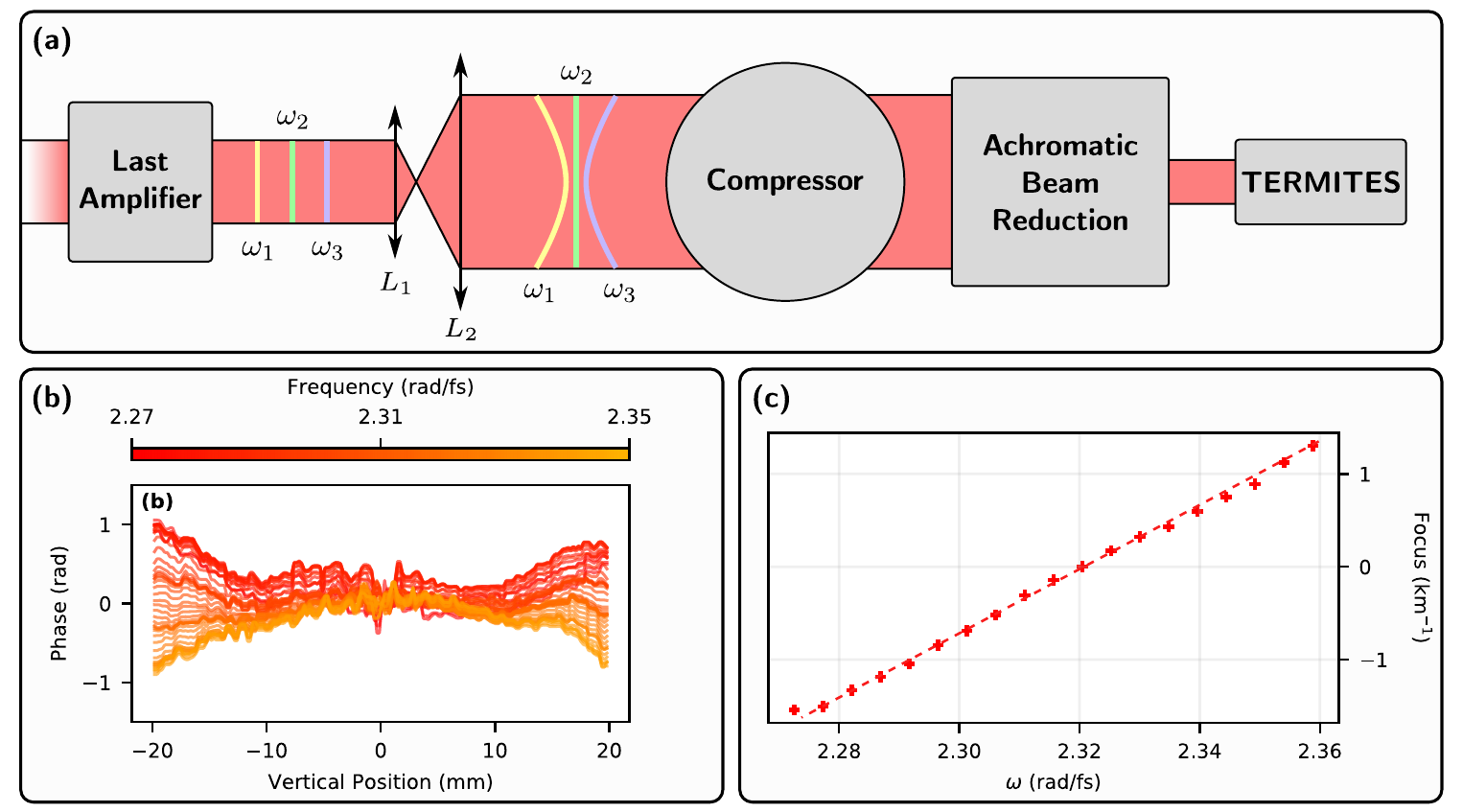}
\caption{\textbf{PFC measurement at LASERIX.} \textbf{(a)}~$L_1$ and $L_2$ are converging singlets forming a magnifying telescope that is used to avoid optical damages in the compressor, but result in a net chromatism.
\textbf{(b)}~Line-outs of the measured spatial phase at different optical frequencies, resolved along the vertical dimension and taken at the center of the beam.
\textbf{(c)}~Second order coefficients of polynomial fit done on the line-outs in panel~(a). The linear fit of these curvatures is shown as a red dashed line and its computed slope is \SI{0.067}{\femto\second/\milli\meter^2}.
}
\label{fig:laserix-measurement-layout}
\end{figure}

The infrared laser beam of LASERIX was characterized in a setup which is illustrated in panel~(a) of  Fig.~\ref{fig:laserix-measurement-layout}. In this laser system, the telescope $L_1$--$L_2$ is placed after the last Ti:Sapphire amplifier to reduce the fluence on compressor optics and avoid optical damage. These large singlet lenses are inherently slightly chromatic, and the large diameter of the beam on the lenses results in an overall significant effect. We performed characterization after the pulse compressor with a TERMITES device, using a telescope composed of achromatic lenses to reduce the diameter of the beam prior to the measurement device. We verified by numerical ray-tracing that this second telescope adds a negligible amount of PFC in the characterized beam~\cite{jeandet20}.

Panel~(b) of Fig.~\ref{fig:laserix-measurement-layout} features vertical lineouts of the spatial phase of the LASERIX beam at different frequencies, as measured by TERMITES. Visually, the phase plots are clearly curved for the outer spectral components ($2.27$ and \SI{2.35}{\radian/\femto\second}), and the curvature evolves smoothly while its sign changes at the central frequency. Of course, curvature is not the only phase component because some other aberrations are visible in the spatial phase, albeit smaller in magnitude. To characterize quantitatively the chromatic curvature, we projected the spectrally-resolved wavefront on Zernike polynomials in order to compute the beam curvature as a function of optical frequency (See Ref.~\cite{jolly20-3} for more details on such a procedure). This data---plotted in panel~(c)---definitively shows that the chromatic curvature observed on panel~(b) generates PFC in the time domain, because the curvature evolves linearly with frequency. For the considered beam diameter of \SI{50}{\milli\meter}, this linear evolution of the curvature corresponds to a delay of \SI{41}{\femto\second} between the center and the edges of the laser beam.

We can confirm the source of this significant spatio-temporal coupling by investigating the properties of the optics in use. As mentioned earlier, PFC is generally added to a laser beam by Keplerian telescopes composed of singlet lenses~\cite{borFemtosecondresolutionPulsefrontDistortion1989,borDistortionFemtosecondLaser1988,heuckChromaticAberrationPetawattclass2006,kempeSpatialTemporalTransformation1992c}. The telescope used in LASERIX is made of plain fused-silica singlets. The PFC induced by such a simple system can be computed using a model obtained by Z. Bor~\cite{borDistortionFemtosecondLaser1988}~(equation~(43)) and does not require any complex ray-tracing. At transverse position~$\mathrm{r}$, the delay $\Delta\tau(r)$ associated to PFC is:

\begin{equation}
\Delta\tau(\mathrm{r}) = \frac{\|\mathrm{r}\|^2}{2\,c\,f_2\,(n-1)}\,\Big(-\lambda_0\frac{\text{d}n}{\text{d}\lambda}\Big)\Big(1+\frac{1}{M}\Big)\,,
\label{eq:bor-pfc-telescope}
\end{equation}

\noindent with $c$ the speed of light, $f_2$ the focal length of lens $L_2$, $n$ and $\frac{\text{d}n}{\text{d}\lambda}$ the refractive parameters of fused-silica~\cite{tanDeterminationRefractiveIndex1998}, $\lambda_0$ the central wavelength of the laser beam, and $M=f_1/f_2$ the telescope magnification. Eq.~\ref{eq:bor-pfc-telescope} predicts delay of \SI{58}{\femto\second} between the center and the edge of the beam, when computed with the parameters of the system. Considering that there might be other sources of weaker PFC in the system, this compares relatively well with the \SI{41}{\femto\second} measured with TERMITES, thus confirming our hypothesis about the main origin of this STC.

\begin{figure}[htb]
\centering
\includegraphics[width=\textwidth]{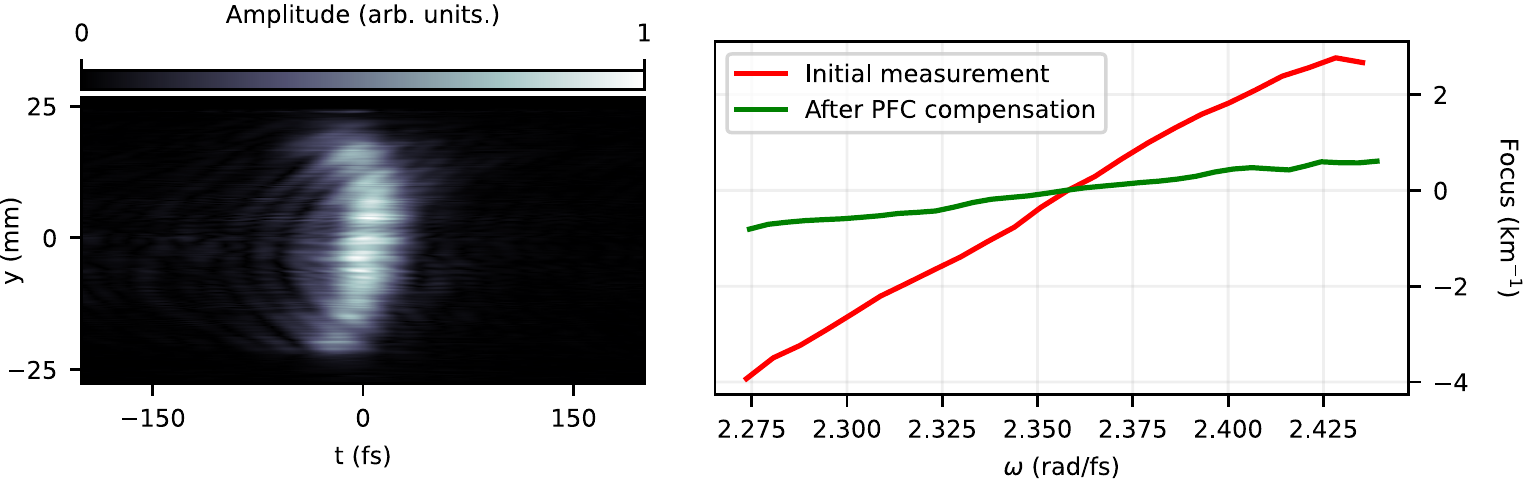}
\captionsetup{format=plain}
\caption{\textbf{Pulse-front curvature on the \emph{Salle Jaune} system.} Slice (left) of the \emph{Salle Jaune} laser spatio-temporal intensity profile, obtained from a TERMITES measurement. The chromatic curvature (right), equivalent to the PFC, was mostly removed when a purposefully designed chromatic doublet was inserted at the end of the laser chain. These two curves are obtained from INSIGHT measurements carried out using the SourceLAB commercial device \cite{Sourcelab}.}
\label{fig:salle-jaune-pfc}
\end{figure}

The consequences of this STC for the LASERIX system are very limited due to the particular applications for this laser. Indeed, the main use so far of the high-energy beam delivered by this system is to create plasmas at the surface of solid targets, which are then used as amplification media for the generation of coherent extreme ultraviolet beams. Because the laser is routinely stretched in time and used in the picosecond regime, the delay of $\simeq$40\,fs that we measured on the edges of the beam is insignificant. Contrary to the case of LASERIX, for laser systems requiring the highest possible intensity, a minimum pulse duration or a designed electric field, even a small amount of PFC could be a significant and unwanted distortion.

PFC was also measured on the \SI{150}{\tera\watt} laser beam of \emph{Salle Jaune}. The left panel of Fig.~\ref{fig:salle-jaune-pfc} shows that the measured pulse front was noticeably curved. The delay value is approximately \SI{40}{\femto\second} at the edge of the beam ($y=\SI{25}{\milli\meter}$), which equals approximately to a PFC of \SI{0.064}{\femto\second/\milli\meter^2}. As this measurement was done with the first prototype of TERMITES, the extracted data is noisy and less reliable than the LASERIX measurement. Nevertheless, the source of this PFC was successfully identified following our measurement. Similarly to the case of LASERIX, it was found that the magnifying telescope before the compressor was responsible for this PFC. Due to the more stringent requirements of the experiments done using Salle Jaune, this aberration was fixed later by inserting a custom compensating afocal doublet~\cite{sainte-marie17,jolly20-1,kabacinski21}, and there are other proposed methods for such compensation~\cite{neauport07,cui19}. The compensation at Salle Jaune is demonstrated in the right panel of Fig.~\ref{fig:salle-jaune-pfc}, where the chromatic curvature has been significantly reduced with the addition of this specifically-designed doublet. Note that most of the measured residual PFC after correction was actually induced by the slightly chromatic lens used to focus the beam into the INSIGHT device.

\section{Effects of Pulse Compression}
\label{sec:spatio-temporal-effects-of-pulse-compression}
One of the main outcomes of the measurement campaigns we carried out on the different ultrashort lasers of Table~1 is that the optical compressor often significantly contributes to STCs on the final output beam. In this section, we will review some of these couplings, starting with the trivial case of Pulse Front Tilt. We will then turn to more advanced couplings induced by compressors, in both phase and amplitude, which are extremely difficult to detect with more traditionnal measurement approaches.

\subsection{Pulse Front Tilt from a Tilted Grating}
\label{subsec:pulse-front-tilt}
Pulse Front Tilt (PFT) is one of the simplest low-order STCs. It is also probably the most common, since it can be induced by any optical system exhibiting angular dispersion (AD), such as a simple prism or grating---but also a pulse stretcher or compressor misaligned in such a way that the AD of the individual components do not exactly compensate each other at the output. In the most common situations encountered with ultrafast lasers, there is actually a one-to-one relationship between PFT and AD~\cite{bor93}, which both describe the same coupling in different spaces (respectively, space-time and space-frequency). When a beam affected by such a PFT/AD is focused, the different frequency components of the beam obviously get focused at different transverse positions, and this then results in transverse spatial chip at focus.

The most common source of PFT/AD on ultrashort lasers is a compressor misalignement where the gratings are not parallel to each other in the horizontal plane. This source of STC is well-known by all users of ultrafast lasers, who typically make sure that it is minimized. This can be achieved e.g. by measuring the residual transverse spatial chirp at the beam focus (for instance using bandpass spectral filters), but the most current practice simply consists in minimizing the residual AD by checking for any elongation of the beam focal spot along the compressor dispersion direction. For this reason, PFT/AD was small---but generally not strictly zero---on most systems we characterized.

\begin{figure}[htb]
\centering
\includegraphics[width=\linewidth]{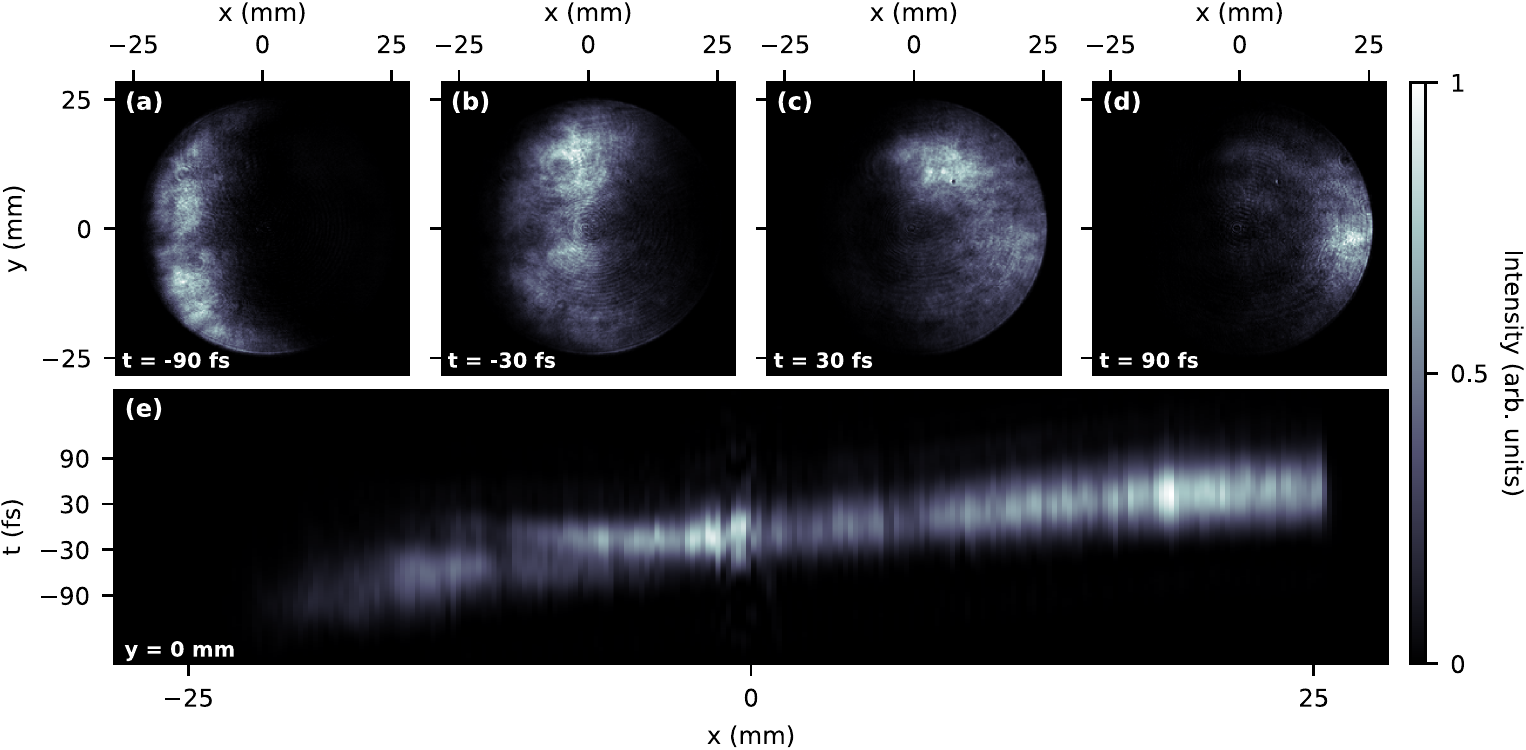}
\caption{\textbf{Measurement of Pulse Front Tilt on the LASERIX laser.}
\mbox{\textbf{(a)--(d)}}~Temporally-resolved spatial intensity profiles of the measured beam, at different times throughout the pulse (see white labels on the images). \textbf{(e)}~Spatio-temporal slice of the beam along the horizontal position. In order to better see the delay added by PFT, we imposed a flat spectral phase to simulate a perfect compression of the pulse in time.}
\label{fig:pft-laserix}
\end{figure}

The situation is different on LASERIX (see Table~1), as mentioned earlier: the particular application of LASERIX does not require the laser pulse to be fully compressed, nor to have a minimized PFT. This is why we observed a significant amount of residual PFT at the compressor output of LASERIX (Fig.~\ref{fig:pft-laserix}), and these measurements provide an instructive direct visualization of this well-known effect. Fig.~\ref{fig:pft-laserix} displays the spatio-temporal intensity profile of LASERIX. PFT manifests itself as a delay on the time of arrival of the pulse, which depends linearly on the transverse position (in the present case, $x$): panels~\mbox{(a)--(d)} show distinctly how the light intensity is located on the left of the beam at early times, and on its right at later times. When such a beam is focused, this temporal streaking of the beam aperture in the near fied leads to an ultrafast rotation of the light propagation direction at focus---an effect called ultrafast wavefront rotation, exploited for instance for the generation of isolated attosecond pulses in the attosecond lighthouse scheme~\cite{vincenti12,wheeler12,quere14}. The measured PFT amounts to \SI{3.6}{\femto\second/\milli\meter}, leading to a total delay of  \SI{180}{\femto\second} over the \SI{50}{\milli\meter} width of the beam. This residual PFT was easily eliminated by rotating one of the compressor's gratings around the vertical axis.

\begin{figure}[htb]
\centering
\includegraphics[width=\linewidth]{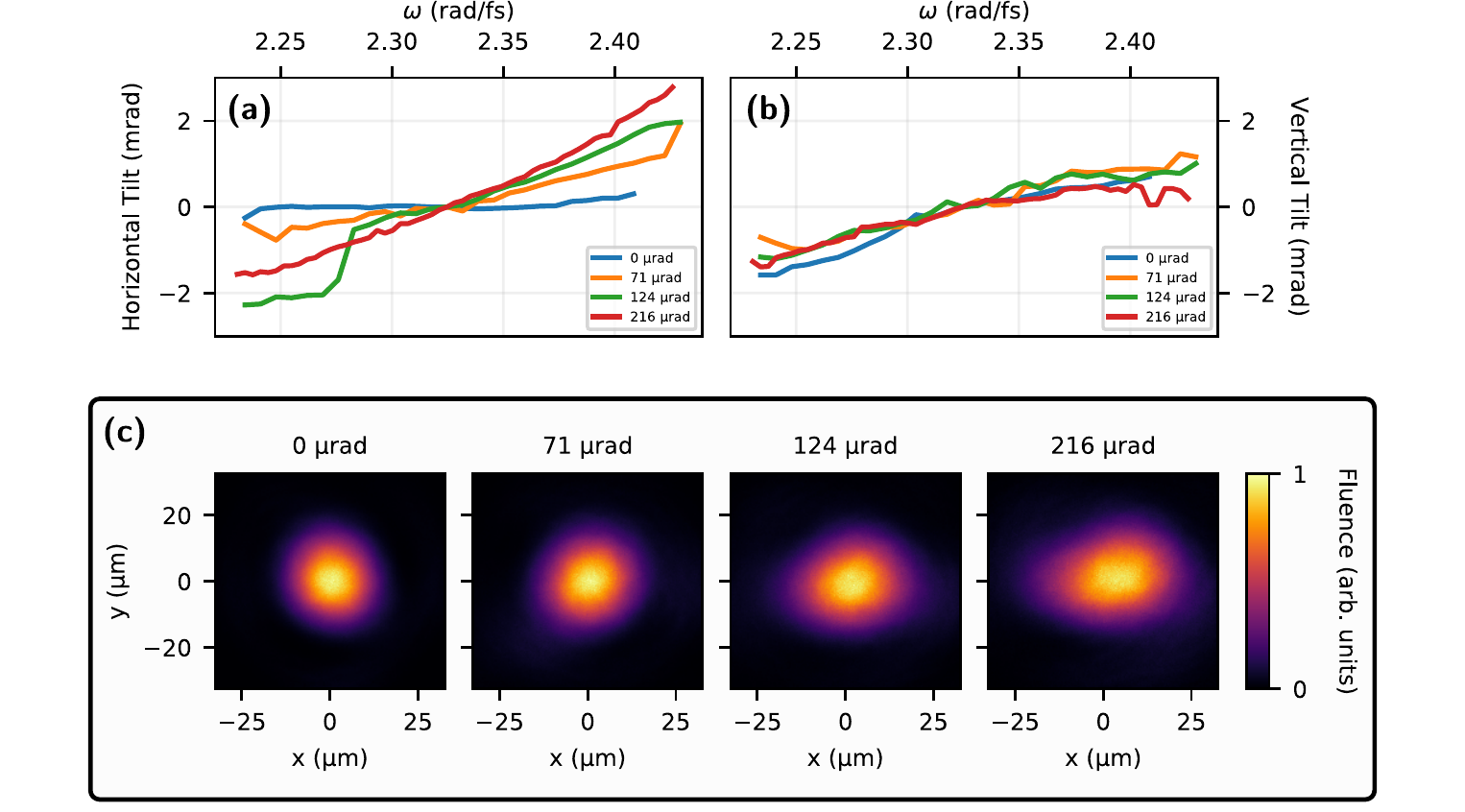}
\caption{\textbf{Fine tuning of the final residual angular dispersion on DRACO TW beamline.}
These measurements were obtained by rotating in small steps one of the compressor gratings in the horizontal direction. Upper panels show the  angular dispersion in the horizontal and vertical directions (wavefront tilts versus frequency), deduced from INSIGHT measurements. Bottom images show the spectrally-integrated profile of the beam obtained at focus with a camera, for each step of the scan. The later is routinely used in most laboratories as a diagnostic for compressor alignement, but has a low accuracy compared to a complete STC characterization: while the horizontal angular dispersion resulting from a 71\,$\mu$rad misalignment of the gratings is clearly detected by INSIGHT, the spectrally-integrated focal spot appears to have no distortion.}
\label{fig:DRACO-AD}
\end{figure}

Measuring this simple coupling obviously does not require methods as advanced as the ones considered here, but diagnostics such as INSIGHT and TERMITES allow for a very accurate determination of the residual PFT, and are thus very helpful for its minimization. The enhanced sensitivity enabled by these techniques compared by the standard minimization routine is illustrated by the results shown in Fig.\ref{fig:DRACO-AD}, obtained from measurements on the DRACO TW laser system. We note that recent results which measured PFT from a misaligned chromatic doublet have also confirmed the high accuracy of complete measurement techniques~\cite{jolly21-2}.

\subsection{Distortions Caused by Defects of Compressor Optics}
\label{subsec:distortions-caused-by-surface-defects-of-compressor-optics}

\begin{figure}[htb]
\centering
\includegraphics[width=\linewidth]{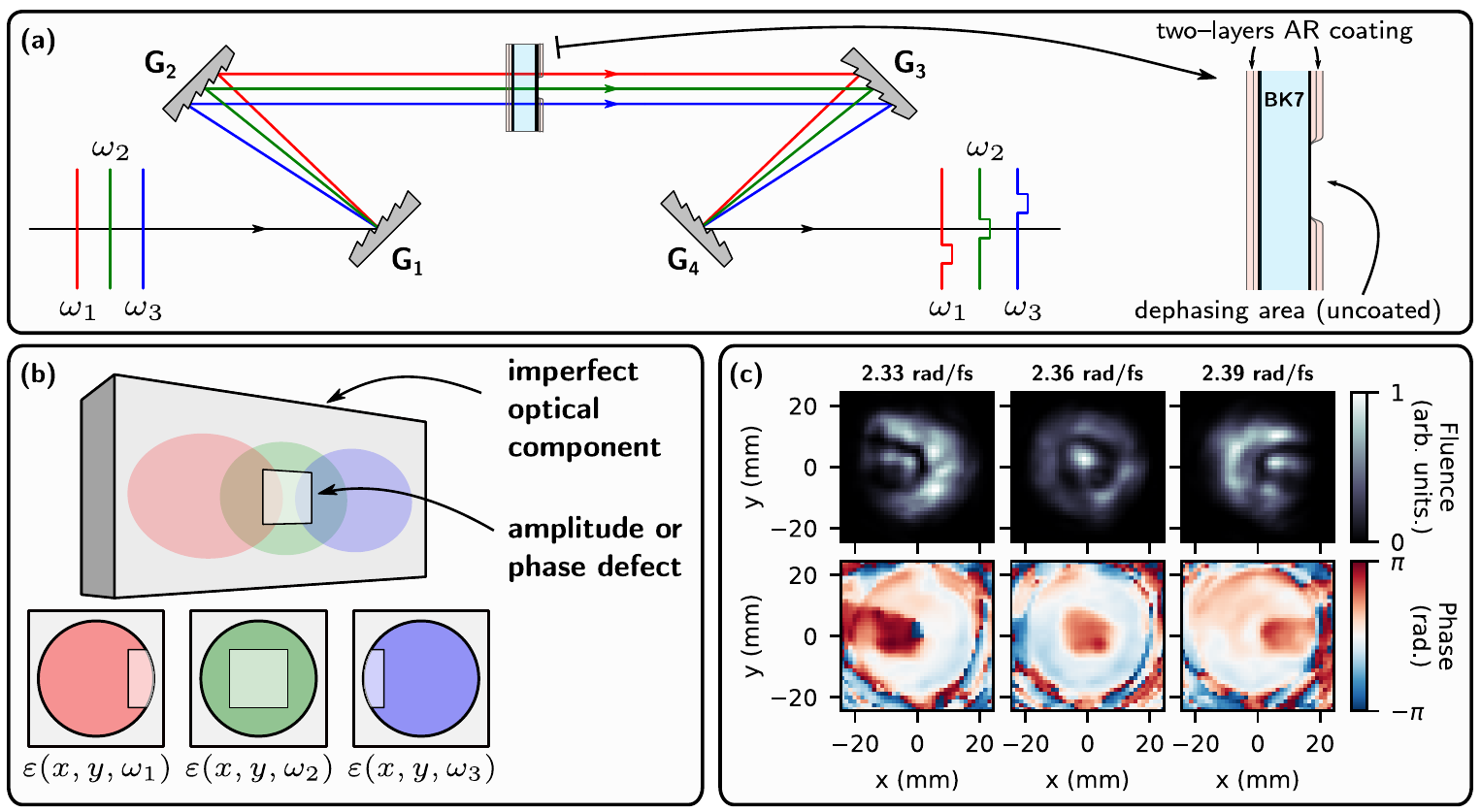}
\caption{\textbf{How compressors can induce chromatic aberrations on ultrashort laser beams}. Panel~(a) displays the typical optical layout of a grating optical compressor (here unfolded for convenience). To demonstrate the effect, we voluntarily placed a custom optical element in the UHI100 laser beam, at a location where it is spatially chirped in the transverse direction. This optical element (glass plate with a partial AR coating, shown on the right) is designed to induce phase aberrations, and to a lesser extent intensity modulations, upon transmission. Panel~(b) illustrates the effect of this custom optical element on the beam within the compressor (uppor line), and then at the compressor output once all frequencies are recombined together (lower line) . Panel~(c) displays the measured spectrally-resolved spatial intensity and phase profiles of the beam at the output of the compressor sketched in (a), at three different frequencies. The induced chromatic aberrations are clearly observed on these experimental results.}
\label{fig:coupling-compressor-explanation}
\end{figure}

Pulse Front Tilt, discussed in the previous subsection, is actually a very peculiar type of aberration, because it is \emph{purely} chromatic: in a beam affected only by PFT, all individual frequencies are totally free from spatial aberrations, but their combination forms an aberrated beam because of the frequency dependence of the wavefront tilt. 
We will now discuss a more complex type of STC where all frequencies are affected by the same type of spatial aberrations (e.g. astigmatism or coma), \emph{and} these aberrations also vary with frequency---i.e., they are chromatic. 

Such couplings can typically be induced in compressors (as well as stretchers, potentially) through the scheme illustrated in Fig.~\ref{fig:coupling-compressor-explanation}(a)--(b): an optical element induces spatial aberrations on the beam, while this beam has a transverse spatial chirp in the compressor (panel~(a)). After the compressor, when the frequency components are again overlapped in space (i.e., spatial chirp has been eliminated), the resulting spatial aberrations then occur at different positions in the beam for different frequencies (panel~(b)). 

Before presenting measurements of such aberrations on different lasers, we illustrate this beam degradation scheme by a simple experiment. A custom-made glass slab was inserted into the compressor of the UHI100 laser, at a location where the beam is spatially chirped. This slab has an anti-reflection (AR) coating on both faces, except for a small 10$\times$10\,mm rectangle on the front face. This local absence of AR coating induces a phase shift of about $\pi$ rad between the part of the beam that propagates through this rectangular section and the rest of the beam, as well as a small amplitude modulation on the spatial profile.

Figure~\ref{fig:coupling-compressor-explanation}(c) shows spectrally-resolved spatial intensity and phase profiles of the beam outcoming from the compressor in such a configuration, measured with INSIGHT. The spatial phase aberration induced on the beam by the tailored glass plate is clearly visible. This aberrations is chromatic: it drifts spatially as frequency changes. The same frequency dependence is observed on the intensity profile. In this case, the modulation originates both from the local variation in plate transmission, and from diffraction resulting from the spatial phase modulation.

In practice, optical elements used inside optical compressors are never perfect---especially as these are large and often state-of-the-art optical elements, such as high-efficiency gratings. As we will now see, this then unavoidably leads to the type of complex STC illustrated by the experiment of Fig.~\ref{fig:coupling-compressor-explanation}. 

\subsubsection{Chromatic phase aberrations}
\begin{figure}[htb]
\centering
\includegraphics[width=\linewidth]{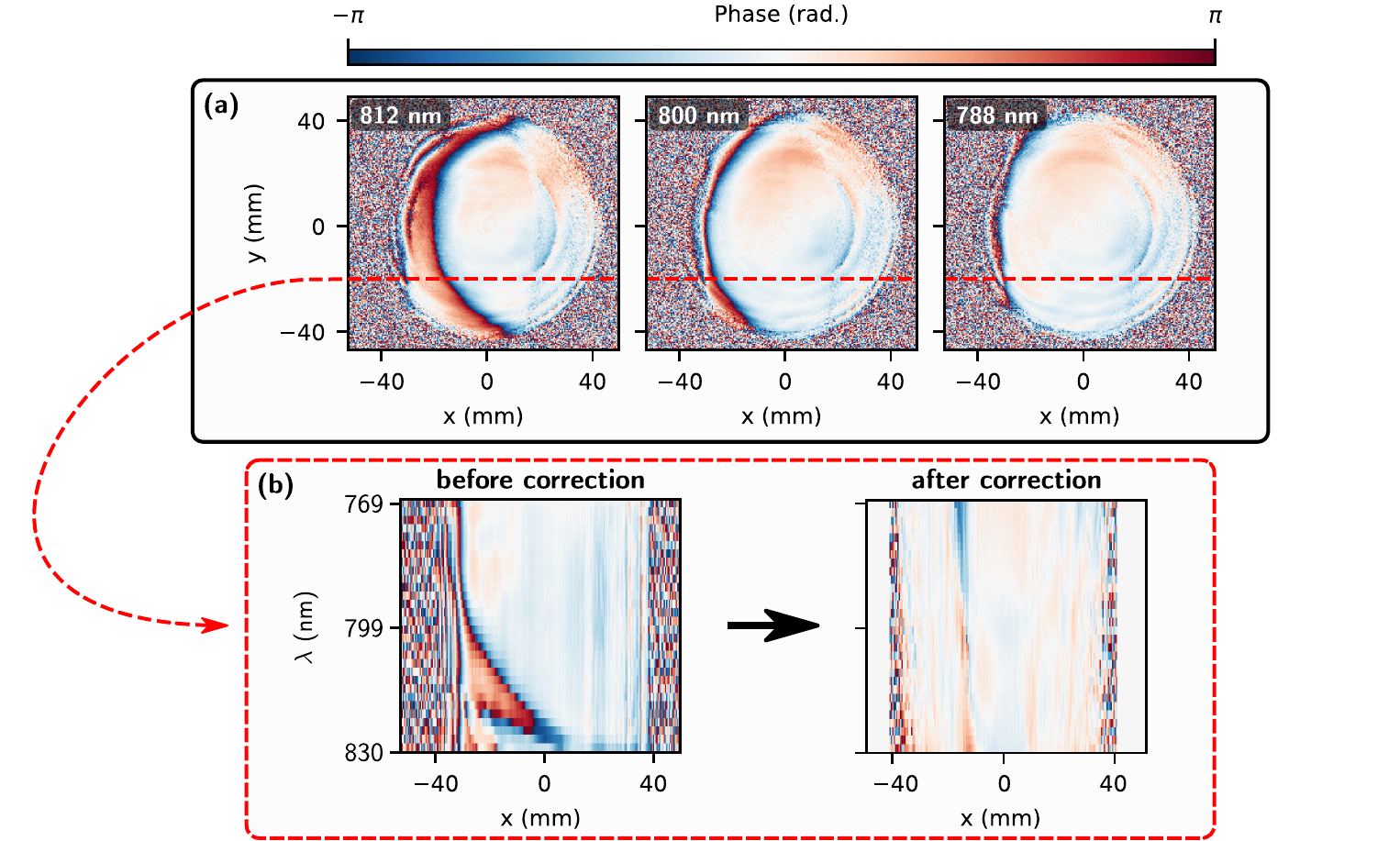}
\caption{\textbf{Chromatic phase aberrations on the UHI100 laser beam.}
\textbf{(a)}~Spectrally-resolved wavefronts of the collimated UHI100 beam at $2.32$, $2.35$ and \SI{2.39}{\radian/\femto\second} (respectively corresponding to wavelengths of $812$, $800$ and \SI{788}{\nano\meter}). \textbf{(b)}~Spatio-spectral phase profile along the horizontal line indicated by the red dashed line in panel~\mbox{(a)}, before (left) and after (right) changing the periscope mirrors in the compressor. Data before correction are from the same measurements as presented in Ref.~\cite{parienteSpaceTimeCharacterization2016} \copyright Nature Publishing, but were processed using an improved algorithm. The small localized defect observed on the phase profile after correction is most likely a measurement artifact.}
\label{fig:uhi100-2015}
\end{figure}

The first complete spatio-temporal characterization of a high-power ultrashort laser beam~\cite{parienteSpaceTimeCharacterization2016} carried out on the UHI100 laser (see Table~1) provided a clear example of a complex chromatic spatial phase aberration, induced by defects of compressor optics through the process illustrated in Fig.~\ref{fig:coupling-compressor-explanation}.
Figure~\ref{fig:uhi100-2015} displays the spectrally-resolved spatial phase profile of UHI100 deduced from this measurement, at three different frequencies. A crescent-shape distortion clearly appears on these phase profiles, and it moves from the right to the left of the beam as frequency increases. This frequency-dependence of the phase defect appears as a 'phase streak' on the spatio-spectral view of panel~(b) (left plot). We emphasize that before this complete spatio-spectral characterization, this very significant phase aberration remained totally undetected with standard (spectrally-integrating) wavefront measurement techniques, due to its highly chromatic character. It was succesfully eliminated (see right plot in Fig.~\ref{fig:uhi100-2015}(b)) by changing the two 20x20 cm mirrors of the compressor periscope that reflect the beam while it is spatially-chirped. This strongly suggests the origin as a faulty surface or coating of at least one of these periscope mirrors, resulting in an effect similar to that in Fig.~\ref{fig:coupling-compressor-explanation}.

\begin{figure}[htb]
\centering
\includegraphics[width=\linewidth]{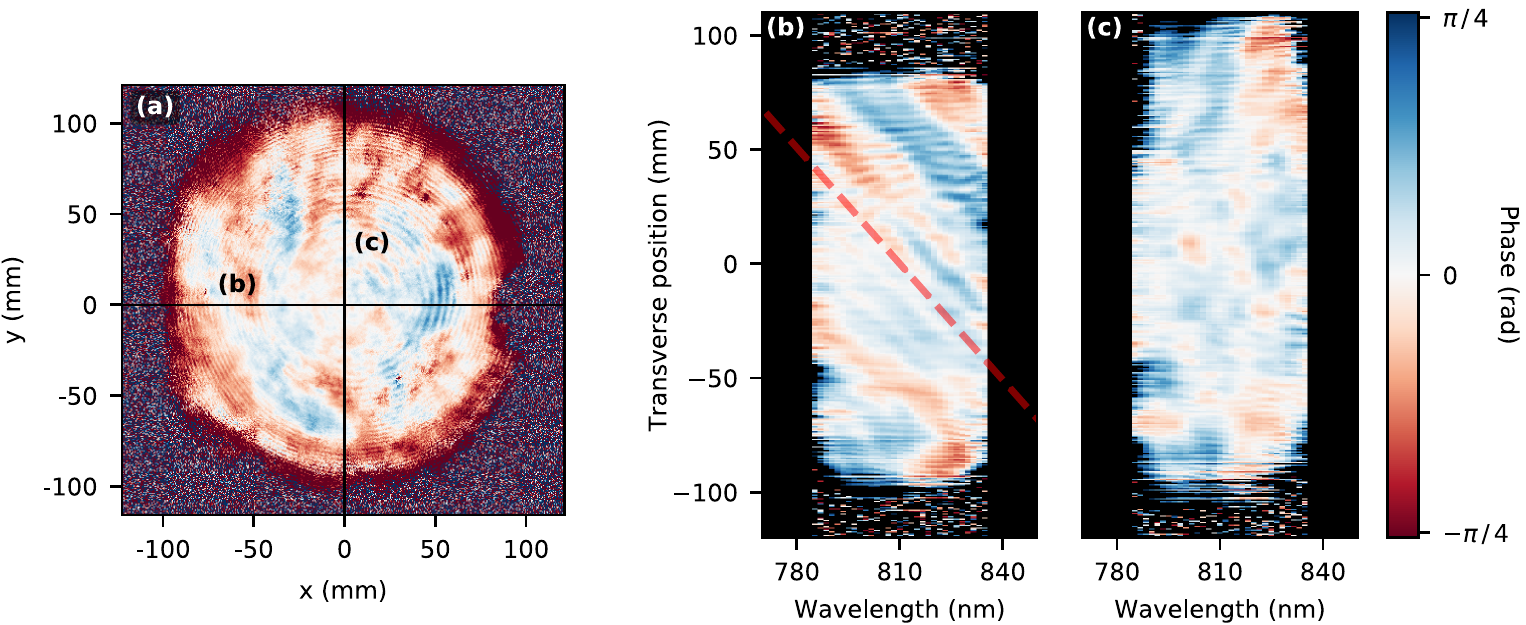}
\caption{\textbf{Spatio-spectral phase of the collimated BELLA beam.}
\textbf{(a)}~BELLA wavefront observed at central wavelength $\lambda_0=\SI{810}{\nano\meter}$.
\textbf{(b)}~Spatio-spectral view of the phase in the $(x, \omega)$~space taken along the horizontal black line in panel~(a).
Oblique red line represents the expected slope of the streak that would be induced by spatial defects of optical elements located within the compressor, given BELLA compressor characteristics.
The origin of this line is arbitrary.
\textbf{(c)}~Spatio-spectral view of the phase in the $(y, \omega)$~space taken along the vertical black line in panel~(a).
The spectrally-averaged wavefront was removed from all panels in order to see exclusively components of the phase that depend on wavelength.  This data is from Ref.~\cite{jeandetSpatiotemporalStructurePetawatt2019} \copyright IOP Publishing.
}
\label{fig:bella-phase-streak}
\vspace{1em}
\end{figure}

A similar 'phase streak' effect was measured on the BELLA petawatt laser beam ~\cite{jeandetSpatiotemporalStructurePetawatt2019}, as shown in Fig.~\ref{fig:bella-phase-streak} (note that some residual ring artefacts can be seen in these results, due to slight delay instabilities that occurred in TERMITES interferometer during the measurement~\cite{jeandet20}). Compared to the crescent-shaped defect previously affecting the UHI-100 beam, the 'phase streak' modulations observed on the spatio-spectral phase of the BELLA beam have a higher spatial frequency, and are distributed more uniformly across the beam profile. They are however much smaller in magnitude ($\approx\pi/4$\,rad. peak-to-valley) than the defect observed on UHI-100. These linear streak patterns are only visible in the horizontal $(x,\omega)$ space (panel~(b)) and not in the $(y,\omega)$ space (panel~(c))---i.e., the frequency-dependent spatial drift exclusively occurs in the dispersion plane of the compressor. These spatio-spectral phase patterns can thus again clearly be attributed to defects of the compressor optical elements, most likely the gratings in this case ~\cite{li17,li18-1,jeandetSpatiotemporalStructurePetawatt2019}. The consequences of such spatio-spectral phase patterns in the near-field on the beam focus and in the temporal domain were also discussed in Ref.~\cite{jeandetSpatiotemporalStructurePetawatt2019}.

\subsubsection{Chromatic effects in intensity}
\label{subsec:spatio-spectral-profile-masking}

Additionally to couplings in phase, compressors (as well as stretchers, in principle) can also induce amplitude couplings on ultrashort laser beams. As is clear from the previous explanations, such couplings can arise whenever the reflectivity of a mirror or the diffraction efficiency of a grating varies spatially, but it can also result from propagation effects initiated by spatial phase modulations induced within the compressor (such as those described in the previous subsection). Finally, a really trivial way to induce such amplitude couplings is when a mechanical or optical element clips the beam while it is spatially chirped in the compressor. Throughout the characterization of different laser systems carried out during this work, we found out that this last case was actually rather common, simply because it can easily be overlooked when standard spatial-only beam diagnostics are used.

\begin{figure}[htb]
\centering
\includegraphics[width=\linewidth]{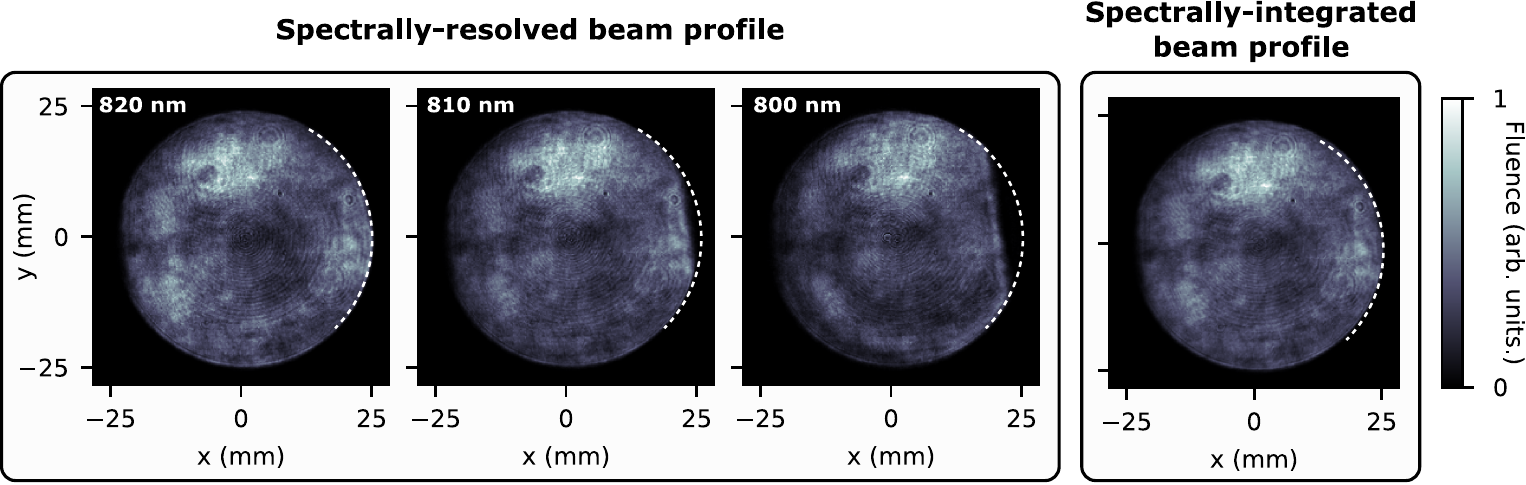}
\caption{\textbf{Spatio-spectral profile masking caused by misalignment of LASERIX compressor.}
The three panels on the left display the spectrally-resolved spatial profile of LASERIX,  at $820$, $810$ and $800$\,nm.
The rightmost panel shows the spatial profile obtained by spectrally integrating these spatio-spectral data. This image could also be obtained using a simple CCD camera.}
\label{fig:laserix-clip}
\end{figure}

Fig.~\ref{fig:laserix-clip} illustrates such a spatio-spectral profile masking, as observed on the LASERIX beam. A vertical shade starts to appear on the right hand side of the beam at \SI{815}{\nano\meter}, and slides into the beam for lower wavelengths, until \SI{800}{\nano\meter} where it masks \SI{5}{\milli\meter} of the beam. The issue is very hard to detect without any spatio-spectral diagnostic tool: as shown in the rightmost panel of Fig.~\ref{fig:laserix-clip}, the spectrally-integrated fluence profile in the near-field, which is typically measured with simple cameras and used as an alignement diagnostic on most lasers, looks normal and would not raise suspicion. In the temporal domain the pulse would be significantly lengthened in the clipped region due to the locally narrower spectrum. The much more complete measurement presented here allowed the LASERIX team to locate the misalignment in the compressor and fix the issue by re-centering the beam on compressor optics.

Similar issues were observed on different lasers, either due to coating degradation on some of the compressor mirrors, or simple clipping by mechanical elements. In all cases, these issues remained undetected with the usual spectrally-integrated diagnostics, but were revealed with the type of measurements presented here. 

\section{Effects of Hollow-Core Fiber Compression}
\label{sec:spatio-temporal-effects-of-hollow-core-fiber-compression}

Hollow-Core Fiber (HCF) compressors are increasingly used at the end of high-power laser systems to further compress ultrashort pulses in time, either to push the pulse duration to the single-cycle regime with established femtosecond lasers up to kHz repetition rates~\cite{chenEfficientHollowFiber2011,boehle14,ouille20}, or to compress pulses from 10-100\,kHz rep-rate technologies from hundreds of femtoseconds or a few picoseconds to tens of femtoseconds~\cite{chen18,nagy19}. This is done by taking advantage of Self Phase Modulation (SPM) which reduces the Fourier-transform limited duration of the pulse by broadening its spectrum. The method essentially relies on focusing pulses in a meter-scale hollow fiber, filled with noble gas at a controlled pressure. Compression in HCF typically results in millijoule-level few-cycle pulses, produced by medium scale facilities such as the \emph{Salle Noire} laser~\cite{ouille20}.

Some spatio-temporal characterization studies of HCF compressed beams~\cite{alonsoCharacterizationSubtwocyclePulses2013,wittingCharacterizationHighintensitySub4fs2011,wittingSub4fsLaserPulse2012} have already been conducted using the SEA-SPIDER and STARFISH techniques, resolving only one transverse spatial dimension. They did not show any strong aberration generated by this non-linear technique under optimal operating conditions. We used TERMITES and INSIGHT to characterize the HCF compression set-up installed in \emph{Salle Noire}, and thus obtained the complete amplitude and phase information on the broadband beam in 3D. We note that for practical experimental reasons, all measurements were carried out right at the fiber output, before the chirp compensation optics used for the final compression of the pulse to few-cycle durations.

\begin{figure}[htb]
\centering
\includegraphics[width=\linewidth]{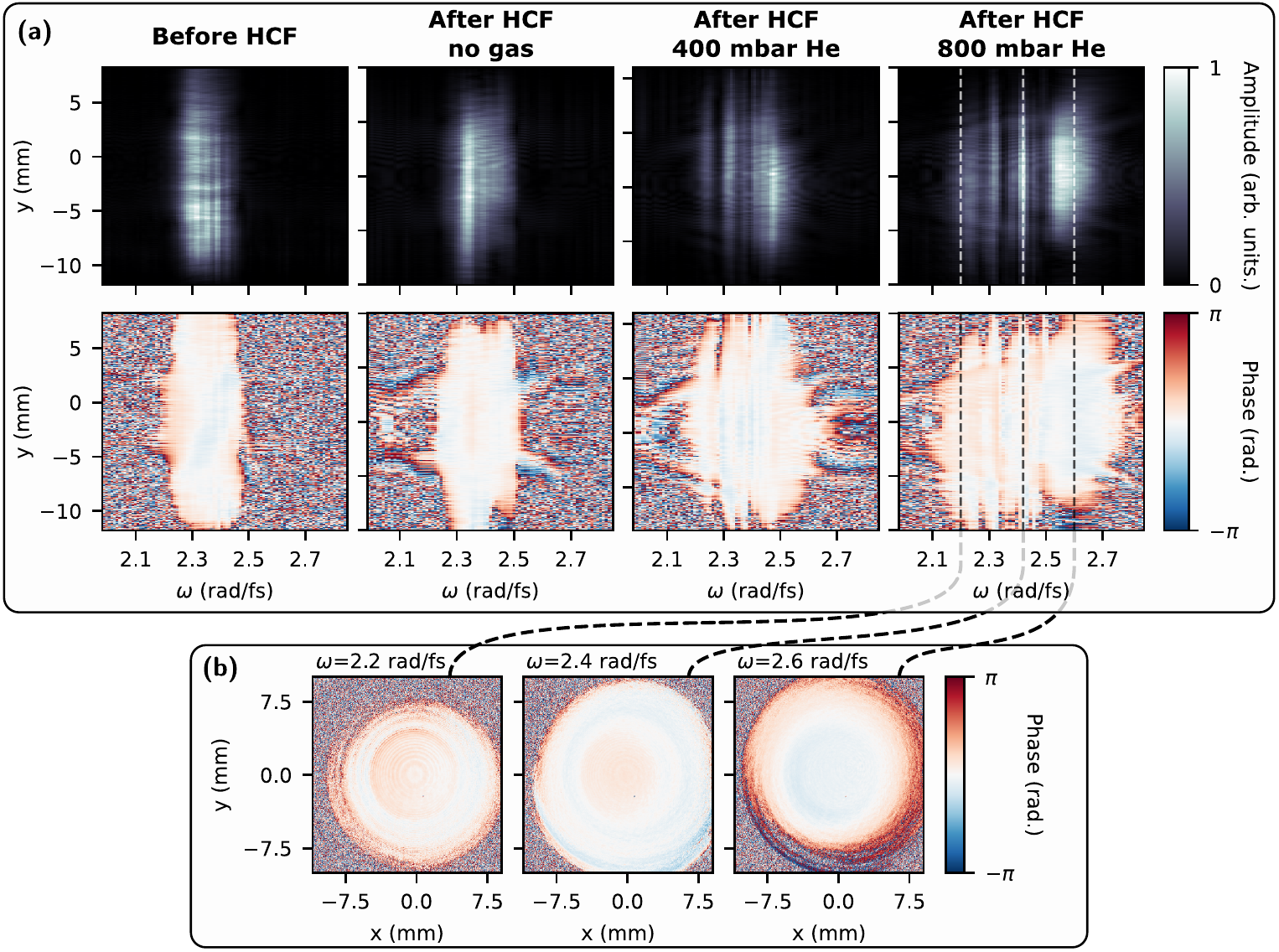}
\caption{\textbf{Spatio-spectral properties of the Salle Noire laser beam, before and after a HCF, for different gas pressures.}
Panel~(a) features spatio-spectral slices of the amplitude and phase profiles mesured by TERMITES. Panel~(b) shows spectrally-resolved spatial wavefronts measured after the fully pressurized fiber, at three particular frequencies.}
\label{fig:hcf-pressure-increase}
\end{figure}

Fig.~\ref{fig:hcf-pressure-increase}(a) shows slices of the spatio-spectral amplitude and phase measured before the HCF, and after the HCF with different levels of gas pressure (and after propagation away from the fiber exit). Amplitude profiles show that the spectral broadening is correctly measured, and that the spectral width significantly increases with helium pressure in the fiber. The phase information is correctly obtained for frequencies where the signal is sufficiently high, and becomes dominated by noise for frequencies with lower intensity. Although the spatio-spectral phase is certainly not perfect, no strong or clearly identified STC can be observed on the beam. Spectral modulations essentialy affect all positions in space identically. The quality of the spatial phase can be seen more clearly in Fig.~\ref{fig:hcf-pressure-increase}(b) where we show 2D spatial phase profiles at three strongly present frequencies on the broadest spectrum (800\,mbar He). This view is enabled by the use of a full 3-D characterization device and shows very clearly the high quality of the phase even for very different frequencies. Note however that we do see a small spatial drift of the spatial profile with frequency (i.e., traverse spatial chirp), most visible in the case with the broadest bandwidth. This small spatial chirp could be due to slight fiber misalignment, leading to the different frequencies leaving the fiber at different angles.

A conclusion that we can draw from these basic measurements is that the technique of using a hollow-core fiber to broaden the spectrum of an ultrashort laser pulse has the potential to be robust and free from spatio-temporal couplings. Indeed, this requires a more investigation to be a general conclusion for all systems, but it is important for CPA laser technology as post-compression techniques are becoming ubiquitous and STCs become more important as the bandwidth becomes so large. This tentative conclusion is consistent with previous amplitude-only or partial measurements made on a range of systems~\cite{weitenberg17,lavenu18,lu18} and full spatio-spectral measurements made on a gas-filled multi-pass cell system~\cite{daher20}.

\section{Conclusion}
In this survey, we have shown that spatio-temporal couplings are present in the multiple stages of
CPA-based laser systems and arise for very different reasons. Spatio-temporal pulse characterization not only allows the optimization of ultrafast light sources, but also offers a deeper understanding of their operation and has the potential to provide new insights about fundamental mechanisms in such cutting-edge laser systems.  

Couplings were sorted depending on the optical component or fundamental process responsible for each of them. We started the survey by confirming that Offner stretchers can introduce significant amounts of spatial chirp even for weak misalignements~(section~\ref{sec:spatio-temporal-effects-of-pulse-stretching},~p.\pageref{sec:spatio-temporal-effects-of-pulse-stretching}). The resulting STCs were observed to be efficicently mitigated by subsequent regenerative amplifiers, at the expense of very significant spectral narrowing of the beam ~(section~\ref{sec:spatio-temporal-effects-of-pulse-amplification},~p.\pageref{sec:spatio-temporal-effects-of-pulse-amplification}). Then, we showed the spatio-spectral effect that can be induced by saturated high-energy amplifiers. We also showed that magnification telescopes based on singlet lenses can induce a noticeable quantity of chromatic curvature, associated to Pulse Front Curvature in the temporal domain~(section~\ref{sec:spatio-temporal-effects-of-beam-imaging},~p.\pageref{sec:spatio-temporal-effects-of-beam-imaging}). Ubiquitous couplings originating from compressors were presented~(section~\ref{sec:spatio-temporal-effects-of-pulse-compression},~p.\pageref{sec:spatio-temporal-effects-of-pulse-compression}), which were either simple STCs caused through misalignment, or complex ones resulting from poor surface quality of constituent optics. Finally, we evaluated the quality of broadband beams generated by Hollow-Core-Fiber post-compression~(section~\ref{sec:spatio-temporal-effects-of-hollow-core-fiber-compression},~p.\pageref{sec:spatio-temporal-effects-of-hollow-core-fiber-compression}). This survey focused on relatively standard chirped-pulse amplification systems, showing the interest of understanding and therefore characterizing the spatio-temporal effects within the amplifier chain rather than solely at the ouput of the system. Our results provide strong motivation to extend such spatio-temporal characterization to a wider range of types of broadband laser systems, both within those systems and at the output, and to link this characterization and understanding to greater number of applications of ultrashort laser pulses.

We conclude this survey with two important points related to ultrafast metrology. First, although some of the observed couplings could have been measured with simpler means, i.e. not necessarily in full 3D, our understanding of the spatio-temporal beam
properties in all the cases was greatly improved through the access to all three dimensions, both in amplitude and phase, which is enabled by the measurement tools used in this work. Second, we emphasize that all measurements presented here were carried out using multi-shots techniques, i.e., requiring many indivual laser shots for a single STC determination. This contrasts with conventional  wisdom in ultrafast metrology, whereby single-shot measurement are preferable. Developing a reliable, accurate and tractable single-shot technique for STC measurements is actually very challenging, although there are some promising developments in that direction~\cite{gabolde06}---yet with rather limited spatial and/or spectral samplings--- that have only recently been applied to high-power systems~\cite{Yoon:21,Kim:21,grace21}. Fortunately, our results show that it is actually not at all a prerequisite to an in-depth analysis of the spatio-temporal properties of ultrashort light sources.

Actually, single-shot STC measurement techniques  will mostly be useful for iterative alignement procedures, in particular of stretchers or compressors, which clearly benefit from very fast measurements of simple and low-order couplings. For more advanced insight into the spatio-temporal properties of ultrashort laser beams, they will probably be truly necessary only in a limited number of very specific cases. The first one is the spatio-temporal characterization of systems with very low repetition rates (i.e., below 1 shot/min)---typically the most powerful laser systems (>1 PW). In such a case, the requirement of a large number of shots for a measurement implies very long acquisition times, which can be very inconvenient or imply a significant risk of a drift in the light source properties during the measurement itself. The second case corresponds to light sources that are intrinsically unstable from shot to shot. This is unlikely to occur for well-controlled laser sources, but could for instance arise once STC measurement techniques are used to investigate non-linear optical effects with a strong sensitivity to the input beam properties. But, in more common cases, our results indicate that STC are stable enough to get a reasonable measurement with multi-shot techniques.

\appendix
\section{APPENDIX: Brief summary of measurement techniques---TERMITES and INSIGHT}
\label{sec:measurement-techniques}

TERMITES and INSIGHT are two techniques for the complete 3D spatio-temporal/spatio-spectral characterization on ultrashort laser beams. They have already been described in details in previous publications~\cite{gallet14,parienteSpaceTimeCharacterization2016,borot18, parienteCaracterisationSpatiotemporelleImpulsions2017,jeandet20}, so we only provide a summary of their principles here. A common point to both techniques is that they rely on spatially-resolved Fourier transform spectroscopy (FTS): a camera is used to resolve the beam along the two spatial coordinates transverse to its propagation direction, while the frequency resolution is provided by a delay scan. The two techniques however strongly differ in the way in which phase information is extracted from the measured data.

TERMITES has been developed independently and simultaneously at CEA-Saclay~\cite{gallet14,parienteSpaceTimeCharacterization2016,parienteCaracterisationSpatiotemporelleImpulsions2017} and Lund~\cite{miranda14}, in slightly different implementations. In both cases, FTS is used to determine an ensemble of cross spectral density functions, and thus compare each point of the unknown beam to a unique reference. In the Saclay version, we compare the spectral phase at any point of the beam, to the spectral phase of a reference taken from its central part. TERMITES can thus be viewed as a self-referenced interferometric technique. It is typically applied to collimated beams, and does not require these beams to be temporally compressed.

In contrast, INSIGHT~\cite{borot18} is typically applied to focused beams---but just like TERMITES it can also work on chirped beams. Spatially-resolved FTS is used to measure the local pulse spectrum at each point of the beam---i.e., using a Michelson interfermeter as a multi-shot hyperspectral camera. From this information, one can direclty determine the spectrally-resolved beam spatial profile. This profile is successively measured in multiple planes (minimum two, more typically three) along the beam propagation direction, close to the beam best focus. Knowing the evolution of the beam intensity profile along this propagation axis for each frequency $\omega$, the spatial phase at each of these frequencies is then determined by applying an iterative phase retrieval algorithm of Gerchberg-Saxton type, separately at each $\omega$.

In this work, measurement campaigns were performed on different lasers with these two techniques, although all lasers could not be systematically be characterized with both techniques. Which technique was used in each case was partly arbitrary, depending on the measurement opportunity we had and on the availability of the devices and trained scientists. We note that the goal of the present work was not a detailed comparison of the two techniques, although this would definitely be of interest. Such a comparison has been done on the BELLA laser~\cite{jeandetSpatiotemporalStructurePetawatt2019}, and we observed that TERMITES was able to resolve spatial modulations of higher frequency than INSIGHT. On the other hand, TERMITES is generally less compact, more difficult to operate, and the data analysis is more challenging. A typical measurement with either INSIGHT or TERMITES can be achieved in typically less than 10 minutes for lasers with a repetition rate of 1 Hz or more. 

While TERMITES typically provides a reconstruction of the complex field $\Tilde{E}(x,y,\omega)$ in the near-field, INSIGHT provides the spatio-spectral field at or around laser focus, i.e., in the far field. However, once the field is known in both amplitude and phase, it can be numerically propagated to any arbitrary plane. Therefore, all results presented in this article are represented in the near-field, whether the measurement was performed with TERMITES or INSIGHT. Once the field $\Tilde{E}(x,y,\omega)$ is known in amplitude and phase, it only requires a Fourier transformation with respect to $\omega$ to calculate the field $E(x,y,t)$~\cite{jolly20-3}. For most of our measurements done with TERMITES or INSIGHT, we also propagated numerically the electric field back to the optical plane of the suspected STC source, which brings additional precision for the analysis. This numerical imaging was performed using plane-wave decomposition, as explained with more details in Ref.~\cite{jeandet20}.

\section{APPENDIX: Numerical simulations of a regenerative amplifier}
\label{sec:simulation-regen}

 We performed simulations of a representative regenerative amplifier with parameters similar to those of the measured amplifier. The simulated regenerative amplifier is built on a Z-shaped cavity folded in-plane thanks to two tilted concave spherical mirrors (radius of curvature 500\,mm) in a confocal geometry. Injection and extraction occur on the same leg using a two-step 20\,mm KDP Pockels cell. The mirror tilt angle is adjusted to 6.15° to compensate\cite{Tache:87} for astigmatism induced by the 15\,mm-long titanium-sapphire laser active medium at 60.4° Brewster angle. This tilted parallel slab is positioned 10\,mm off-focus along the cavity axis, close to the actual configuration of the measured amplifier. Titanium-doped sapphire is considered as a four-level laser medium with anisotropic wavelength-dependent emission cross-sections modelled by a vibronic Poisson distribution\cite{Eggleston:88}. Kramers-Kronig spectral phase is deduced by MacLaurin integration\cite{Ohta:88}. It adds to undoped-material dispersion effects accounted for by undoped sapphire Sellmeier equations\cite{Malitson:62}. Refraction through interfaces, reflexion on mirrors, and beam propagation in anisotropic KDP and sapphire relies on a vectorial non-paraxial Fourier method\cite{McLeod:14}. Broadband pulse amplification with gain saturation is numerically calculated by a Fourier split-step method similar to\cite{Morice:1999,Paschotta:17} with adaptive step-size and adaptive time slices.

A typical energy of 1\,nJ is injected in the cavity with a near-field described in the $x,y,\omega$ space by a standard Gaussian profile with a spatial chirp in the $x$-direction and with a central wavelength of $\lambda_0=2\pi/\omega_0=793$\,nm. A quadratic temporal phase $\phi_2=3.723$\,rad/ps$^{2}$ is applied, corresponding to the stretch of a 21.5\,fs Fourier-transformed pulse to $\tau=35$\,ps full-width at half-maximum (the stretched pulse duration is scaled-down by one order of magnitude compared to experimental value to speed-up calculations). Nonlinear effects related to peak power such as optical Kerr effects and two-photon absorption are therefore turned off in these simulations. Un-aberrated Gaussian beam waist radius $w_0=\mathrm{FWHM}/\sqrt{2\log2}$ is adjusted to match cavity-mode diameter $\mathrm{FWHM} = 0.360$\,mm. The collimated, spatially-chirped field is Fourier-transformed in time and imaged on the cavity first flat mirror. The small-signal gain is adjusted through excited-state fraction of Ti$^{3+}$ ions to saturate amplification after 12 to 15 round-trips (24 to 30 passes) while extracting a typical energy of 2\,mJ. 

\section*{Funding}

The research leading to this work has received funding from the European Union's Horizon 2020 research and innovation programme (ExCoMet ERC grant agreement no. 694596, TERMITES ERC grant agreement no. 727427, LASERLAB grant agreement no. 871124, CREMLINplus grant agreement 871072), by Investissements d’Avenir LabEx PALM (ANR-10-LABX-0039-PALM, project EXYT), by the CEA’s DRF Impulsion program, and has received financial support from the Amplitude Laser Group. The work at the BELLA facility was supported by the Director, Office of Science, Office of High Energy Physics of the U.S. Department of Energy under Contract No. DE-AC02-05CH11231.

\section*{Acknowledgements}

The authors would like to thank Csaba Tóth, and Hann-Shin Mao for help with the BELLA measurements, Fabrice Reau and David Garzella for operating the UHI-100 laser system, as well as Elsa Baynard and Alok Kumar Pandey for assisting measurements on the LASERIX laser system.

\section*{Disclosures}

The authors declare no conflicts of interest.

\section*{Data Availability}

Data underlying the results presented in this paper are not publicly available at this time but may be obtained from the authors upon reasonable request.

\bibliography{article-characterization-survey}

\begin{thebibliography}{10}
\newcommand{\enquote}[1]{``#1''}

\bibitem{akturk10}
S.~Akturk, X.~Gu, P.~Bowlan, and R.~Trebino, \enquote{Spatio-temporal couplings
  in ultrashort laser pulses,} {\protect\JournalTitle{Journal of Optics}}
  \textbf{12}, 093001 (2010).

\bibitem{ARCO}
{{Amplitude Laser}}, \enquote{Arco,}
  \url{https://amplitude-laser.com/products/lasers-for-science/arco/}. [Online;
  accessed 14-April-2021].

\bibitem{ouille20}
M.~Ouillé, A.~Vernier, F.~B{\"o}hle, M.~Bocoum, A.~Jullien, M.~Lozano, J.-P.
  Rousseau, Z.~Cheng, D.~Gustas, A.~Blumenstein, P.~Simon, S.~Haessler,
  J.~Faure, T.~Nagy, and R.~Lopez-Martens, \enquote{Relativistic-intensity
  near-single-cycle light waveforms at khz repetition rate,}
  {\protect\JournalTitle{Light: Science \& Applications}} \textbf{9}, 47
  (2020).

\bibitem{Guilbaud:15}
O.~Guilbaud, G.~V. Cojocaru, L.~Li, O.~Delmas, R.~G. Ungureanu, R.~A. Banici,
  S.~Kazamias, K.~Cassou, O.~Neveu, J.~Demailly, E.~Baynard, M.~Pittman, A.~L.
  Marec, A.~Klisnick, P.~Zeitoun, D.~Ursescu, and D.~Ros, \enquote{Gain
  dynamics in quickly ionized plasma for seeded operated soft x-ray lasers,}
  {\protect\JournalTitle{Opt. Lett.}} \textbf{40}, 4775--4778 (2015).

\bibitem{Schramm_2017}
U.~Schramm, M.~Bussmann, A.~Irman, M.~Siebold, K.~Zeil, D.~Albach, C.~Bernert,
  S.~Bock, F.~Brack, J.~Branco, J.~Couperus, T.~Cowan, A.~Debus, C.~Eisenmann,
  M.~Garten, R.~Gebhardt, S.~Grams, U.~Helbig, A.~Huebl, T.~Kluge, A.~Köhler,
  J.~Krämer, S.~Kraft, F.~Kroll, M.~Kuntzsch, U.~Lehnert, M.~Loeser,
  J.~Metzkes, P.~Michel, L.~Obst, R.~Pausch, M.~Rehwald, R.~Sauerbrey,
  H.~Schlenvoigt, K.~Steiniger, and O.~Zarini, \enquote{First results with the
  novel petawatt laser acceleration facility in dresden,}
  {\protect\JournalTitle{Journal of Physics: Conference Series}} \textbf{874},
  012028 (2017).

\bibitem{nakamura17}
K.~Nakamura, H.-S. Mao, A.~J. Gonsalves, H.~Vincenti, D.~E. Mittelberger,
  J.~Daniels, A.~Magana, C.~Toth, and W.~P. Leemans, \enquote{Diagnostics,
  control and performance parameters for the {BELLA} high repetition rate
  petawatt class laser,} {\protect\JournalTitle{IEEE Journal of Quantum
  Electronics}} \textbf{53} (2017).

\bibitem{gabolde06}
P.~Gabolde and R.~Trebino, \enquote{Single-shot measurement of the full
  spatiotemporal field of ultrashort pulses with multispectral digital
  holography,} {\protect\JournalTitle{Optics Express}} \textbf{14},
  11460--11467 (2006).

\bibitem{miranda14}
M.~Miranda, M.~Kotur, P.~Rudawski, C.~Guo, A.~Harth, A.~L’Huillier, and C.~L.
  Arnold, \enquote{Spatiotemporal characterization of ultrashort laser pulses
  using spatially resolved fourier transform spectrometry,}
  {\protect\JournalTitle{Optics Letters}} \textbf{39}, 5142--5144 (2014).

\bibitem{parienteSpaceTimeCharacterization2016}
G.~Pariente, V.~Gallet, A.~Borot, O.~Gobert, and F.~Quéré,
  \enquote{Space\textendash{}time characterization of ultra-intense femtosecond
  laser beams,} {\protect\JournalTitle{Nature Photonics}} \textbf{10}, 547--553
  (2016).

\bibitem{borot18}
A.~Borot and F.~Quéré, \enquote{Spatio-spectral metrology at focus of
  ultrashort lasers: a phase-retrieval approach,} {\protect\JournalTitle{Optics
  Express}} \textbf{26}, 26444--26461 (2018).

\bibitem{dorrer19}
C.~Dorrer, \enquote{Spatiotemporal metrology of broadband optical pulses,}
  {\protect\JournalTitle{IEEE Journal of Selected Topics in Quantum
  Electronics}} \textbf{25}, 3100216 (2019).

\bibitem{jolly20-3}
S.~W. Jolly, O.~Gobert, and F.~Quéré, \enquote{Spatio-temporal
  characterization of ultrashort laser beams: a tutorial,}
  {\protect\JournalTitle{Journal of Optics}} \textbf{22}, 103501 (2020).

\bibitem{Sourcelab}
Sourcelab, \enquote{Insight, spatio-temporal metrology at focus of ultra-short
  laser pulses,}
  \url{https://www.sourcelab-plasma.com/laser-shaping/beam-shaping-catalog/insight/}.
  [Online; accessed 14-April-2021].

\bibitem{bromage2019}
J.~Bromage, S.-W. Bahk, I.~A. Begishev, C.~Dorrer, M.~J. Guardalben, B.~N.
  Hoffman, J.~Oliver, R.~G. Roides, E.~M. Schiesser, M.~J. Shoup~III, and
  et~al., \enquote{Technology development for ultraintense all-opcpa systems,}
  {\protect\JournalTitle{High Power Laser Science and Engineering}} \textbf{7},
  e4 (2019).

\bibitem{schmidt14}
B.~E. Schmidt, N.~Thiré, M.~Boivin, A.~Laramée, F.~Poitras, G.~Lebrun,
  T.~Ozaki, H.~Ibrahim, and F.~Légaré, \enquote{Frequency domain optical
  parametric amplification,} {\protect\JournalTitle{Nature Communications}}
  \textbf{5}, 3643 (2014).

\bibitem{fsaifes20}
I.~Fsaifes, L.~Daniault, S.~Bellanger, M.~Veinhard, J.~Bourderionnet, C.~Larat,
  E.~Lallier, E.~Durand, A.~Brignon, and J.-C. Chanteloup, \enquote{Coherent
  beam combining of 61 femtosecond fiber amplifiers,}
  {\protect\JournalTitle{Optics Express}} \textbf{28}, 20152--20161 (2020).

\bibitem{stark21}
H.~Stark, J.~Buldt, M.~M\"{u}ller, A.~Klenke, and J.~Limpert, \enquote{1 kw, 10
  mj, 120 fs coherently combined fiber cpa laser system,}
  {\protect\JournalTitle{Optics Letters}} \textbf{46}, 969--972 (2021).

\bibitem{kessler04}
T.~J. Kessler, J.~Bunkenburg, H.~Huang, A.~Kozlov, and D.~D. Meyerhofer,
  \enquote{Demonstration of coherent addition of multiple gratings for
  high-energy chirped-pulse-amplified lasers,} {\protect\JournalTitle{Optics
  Letters}} \textbf{29}, 635--637 (2004).

\bibitem{li15}
Z.~Li, S.~Li, C.~Wang, Y.~Xu, F.~Wu, Y.~Li, and Y.~Leng, \enquote{Stable and
  near fourier-transform-limit 30fs pulse compression with a tiled grating
  compressor scheme,} {\protect\JournalTitle{Optics Express}} \textbf{23},
  33386--33395 (2015).

\bibitem{liu20}
J.~Liu, X.~Shen, Z.~Si, C.~Wang, C.~Zhao, X.~Liang, Y.~Leng, and R.~Li,
  \enquote{In-house beam-splitting pulse compressor for high-energy petawatt
  lasers,} {\protect\JournalTitle{Optics Express}} \textbf{28}, 22978--22991
  (2020).

\bibitem{li21}
Z.~Li, Y.~Kato, and J.~Kawanaka, \enquote{Simulating an ultra-broadband concept
  for exawatt-class lasers,} {\protect\JournalTitle{Scientific Reports}}
  \textbf{11}, 151 (2021).

\bibitem{cheriauxAberrationfreeStretcherDesign1996}
G.~Cheriaux, B.~Walker, L.~F. Dimauro, P.~Rousseau, F.~Salin, and J.~P.
  Chambaret, \enquote{Aberration-free stretcher design for ultrashort-pulse
  amplification,} {\protect\JournalTitle{Optics Letters}} \textbf{21}, 414
  (1996).

\bibitem{jeandetSpatiotemporalStructurePetawatt2019}
A.~Jeandet, A.~Borot, K.~Nakamura, S.~W. Jolly, A.~J. Gonsalves, C.~Toth, H.-S.
  Mao, W.~P. Leemans, and F.~Quéré, \enquote{Spatio-temporal structure of a
  petawatt femtosecond laser beam,} {\protect\JournalTitle{Journal of Physics:
  Photonics}}  (2019).

\bibitem{frantzTheoryPulsePropagation1963}
L.~M. Frantz and J.~S. Nodvik, \enquote{Theory of {{Pulse Propagation}} in a
  {{Laser Amplifier}},} {\protect\JournalTitle{Journal of Applied Physics}}
  \textbf{34}, 2346--2349 (1963).

\bibitem{durfeeDesignImplementationTWclass1998}
C.~Durfee, S.~Backus, M.~Murnane, and H.~Kapteyn, \enquote{Design and
  implementation of a {{TW}}-class high-average power laser system,}
  {\protect\JournalTitle{IEEE Journal of Selected Topics in Quantum
  Electronics}} \textbf{4}, 395--406 (1998).

\bibitem{giambrunoDesign10PW2011}
F.~Giambruno, C.~Radier, G.~Rey, and G.~Chériaux, \enquote{Design of a 10
  {{PW}} (150 {{J}}/15 fs) peak power laser system with {{Ti}}:sapphire medium
  through spectral control,} {\protect\JournalTitle{Applied Optics}}
  \textbf{50}, 2617 (2011).

\bibitem{borDistortionFemtosecondLaser1988}
Z.~Bor, \enquote{Distortion of {{Femtosecond Laser Pulses}} in {{Lenses}} and
  {{Lens Systems}},} {\protect\JournalTitle{Journal of Modern Optics}}
  \textbf{35}, 1907--1918 (1988).

\bibitem{borDistortionFemtosecondLaser1989}
Z.~Bor, \enquote{Distortion of femtosecond laser pulses in lenses,}
  {\protect\JournalTitle{Optics Letters}} \textbf{14}, 119 (1989).

\bibitem{jeandet20}
A.~Jeandet, \enquote{{Spatio-temporal characterization of femtosecond laser
  pulses using self-referenced Fourier transform spectroscopy},} Ph.D. thesis,
  Université Paris-Saclay (2020).
  \url{https://tel.archives-ouvertes.fr/tel-03018886/document}.

\bibitem{borFemtosecondresolutionPulsefrontDistortion1989}
Z.~Bor, Z.~Gogolak, and G.~Szabo, \enquote{Femtosecond-resolution pulse-front
  distortion measurement by time-of-flight interferometry,}
  {\protect\JournalTitle{Optics letters}} \textbf{14}, 862--864 (1989).

\bibitem{heuckChromaticAberrationPetawattclass2006}
H.-M. Heuck, P.~Neumayer, T.~K{\"u}hl, and U.~Wittrock, \enquote{Chromatic
  aberration in petawatt-class lasers,} {\protect\JournalTitle{Applied Physics
  B}} \textbf{84}, 421--428 (2006).

\bibitem{kempeSpatialTemporalTransformation1992c}
M.~Kempe, U.~Stamm, B.~Wilhelmi, and W.~Rudolph, \enquote{Spatial and temporal
  transformation of femtosecond laser pulses by lenses and lens systems,}
  {\protect\JournalTitle{Journal of the Optical Society of America B}}
  \textbf{9}, 1158 (1992).

\bibitem{tanDeterminationRefractiveIndex1998}
C.~Tan, \enquote{Determination of refractive index of silica glass for infrared
  wavelengths by {{IR}} spectroscopy,} {\protect\JournalTitle{Journal of
  Non-Crystalline Solids}} \textbf{223}, 158--163 (1998).

\bibitem{sainte-marie17}
A.~Sainte-Marie, O.~Gobert, and F.~Quéré, \enquote{Controlling the velocity
  of ultrashort light pulses in vacuum through spatio-temporal couplings,}
  {\protect\JournalTitle{Optica}} \textbf{4}, 1298--1304 (2017).

\bibitem{jolly20-1}
S.~W. Jolly, O.~Gobert, A.~Jeandet, and F.~Quéré, \enquote{Controlling the
  velocity of a femtosecond laser pulse using refractive lenses,}
  {\protect\JournalTitle{Optics Express}} \textbf{28}, 4888--4897 (2020).

\bibitem{kabacinski21}
A.~Kabacinski, K.~Oubrerie, J.-P. Goddet, J.~Gautier, F.~Tissandier,
  O.~Kononenko, A.~Tafzi, A.~Leblanc, S.~Sebban, and C.~Thaury,
  \enquote{Measurement and control of main spatio-temporal couplings in a {CPA}
  laser chain,} {\protect\JournalTitle{Journal of Optics}} \textbf{23}, 06LT01
  (2021).

\bibitem{neauport07}
J.~N{\'e}auport, N.~Blanchot, C.~Rouyer, and C.~Sauteret, \enquote{Chromatism
  compensation of the {PETAL} multipetawatt high-energy laser,}
  {\protect\JournalTitle{Applied Optics}} \textbf{46}, 1568--1574 (2007).

\bibitem{cui19}
Z.~Cui, J.~Kang, A.~Guo, H.~Zhu, Q.~Yang, P.~Zhu, M.~Sun, Q.~Gao, D.~Liu,
  X.~Ouyang, Z.~Zhang, H.~Wei, X.~Liang, C.~Zhang, S.~Yang, D.~Zhang, X.~Xie,
  and J.~Zhu, \enquote{Dynamic chromatic aberration pre-compensation scheme for
  ultrashort petawatt laser systems,} {\protect\JournalTitle{Optics Express}}
  \textbf{27}, 16812--16822 (2019).

\bibitem{bor93}
Z.~Bor, B.~R{\'a}cz, G.~Szab{\'o}, M.~Hilbert, and H.~A. Hazim,
  \enquote{Femtosecond pulse front tilt caused by angular dispersion,}
  {\protect\JournalTitle{Optical Engineering}} \textbf{32}, 2501--2504 (1993).

\bibitem{vincenti12}
H.~Vincenti and F.~Quéré, \enquote{Attosecond lighthouses: How to use
  spatiotemporally coupled light fields to generate isolated attosecond
  pulses,} {\protect\JournalTitle{Physical Review Letters}} \textbf{108},
  113904 (2012).

\bibitem{wheeler12}
J.~A. Wheeler, A.~Borot, S.~Monchoce, H.~Vincenti, A.~Ricci, A.~Malvache,
  R.~Lopez-Martens, and F.~Quéré, \enquote{Attosecond lighthouses from plasma
  mirrors,} {\protect\JournalTitle{Nature Photonics}} \textbf{6}, 829--833
  (2012).

\bibitem{quere14}
F.~Quéré, H.~Vincenti, A.~Borot, S.~Monchocé, T.~J. Hammond, K.~T. Kim,
  J.~A. Wheeler, C.~Zhang, T.~Ruchon, T.~Auguste, J.~F. Hergott, D.~M.
  Villeneuve, P.~B. Corkum, and R.~Lopez-Martens, \enquote{Applications of
  ultrafast wavefront rotation in highly nonlinear optics,}
  {\protect\JournalTitle{Journal of Physics B: Atomic, Molecular and Optical
  Physics}} \textbf{47}, 124004 (2014).

\bibitem{jolly21-2}
S.~W. Jolly, O.~Gobert, and F.~Qu{\'e}r{\'e}, \enquote{Spatio-spectral
  characterization of ultrashort laser pulses with a birefringent delay line,}
  {\protect\JournalTitle{OSA Continuum}} \textbf{4}, 2044--2051 (2021).

\bibitem{li17}
Z.~Li, K.~Tsubakimoto, H.~Yoshida, Y.~Nakata, and N.~Miyanaga,
  \enquote{Degradation of femtosecond petawatt laser beams:
  {Spatio}-temporal/spectral coupling induced by wavefront errors of
  compression gratings,} {\protect\JournalTitle{Applied Physics Express}}
  \textbf{10}, 102702 (2017).

\bibitem{li18-1}
Z.~Li and N.~Miyanaga, \enquote{Simulating ultra-intense femtosecond lasers in
  the 3-dimensional space-time domain,} {\protect\JournalTitle{Optics Express}}
  \textbf{26}, 8453--8469 (2018).

\bibitem{chenEfficientHollowFiber2011}
X.~Chen, A.~Malvache, A.~Ricci, A.~Jullien, and R.~{Lopez-Martens},
  \enquote{Efficient hollow fiber compression scheme for generating
  multi-{{mJ}}, carrier-envelope phase stable, sub-5 fs pulses,}
  {\protect\JournalTitle{Laser Physics}} \textbf{21}, 198--201 (2011).

\bibitem{boehle14}
F.~Böhle, M.~Kretschmar, A.~Jullien, M.~Kovacs, M.~Miranda, R.~Romero,
  H.~Crespo, U.~Morgner, P.~Simon, R.~Lopez-Martens, and T.~Nagy,
  \enquote{Compression of {CEP}-stable multi-{mJ} laser pulses down to 4\,fs in
  long hollow fibers,} {\protect\JournalTitle{Laser Physics Letters}}
  \textbf{11}, 095401 (2014).

\bibitem{chen18}
B.-H. Chen, M.~Kretschmar, D.~Ehberger, A.~Blumenstein, P.~Simon, P.~Baum, and
  T.~Nagy, \enquote{Compression of picosecond pulses from a thin-disk laser to
  30fs at 4w average power,} {\protect\JournalTitle{Optics Express}}
  \textbf{26}, 3861--3869 (2018).

\bibitem{nagy19}
T.~Nagy, S.~H\"{a}drich, P.~Simon, A.~Blumenstein, N.~Walther, R.~Klas,
  J.~Buldt, H.~Stark, S.~Breitkopf, P.~J\'{o}j\'{a}rt, I.~Seres,
  Z.~V\'{a}rallyay, T.~Eidam, and J.~Limpert, \enquote{Generation of
  three-cycle multi-millijoule laser pulses at 318 w average power,}
  {\protect\JournalTitle{Optica}} \textbf{6}, 1423--1424 (2019).

\bibitem{alonsoCharacterizationSubtwocyclePulses2013}
B.~Alonso, M.~Miranda, F.~Silva, V.~Pervak, J.~Rauschenberger,
  J.~San~Rom{\'a}n, {\'I}.~J. Sola, and H.~Crespo, \enquote{Characterization of
  sub-two-cycle pulses from a hollow-core fiber compressor in the
  spatiotemporal and spatiospectral domains,} {\protect\JournalTitle{Applied
  Physics B}} \textbf{112}, 105--114 (2013).

\bibitem{wittingCharacterizationHighintensitySub4fs2011}
T.~Witting, F.~Frank, C.~A. Arrell, W.~A. Okell, J.~P. Marangos, and J.~W.~G.
  Tisch, \enquote{Characterization of high-intensity sub-4-fs laser pulses
  using spatially encoded spectral shearing interferometry,}
  {\protect\JournalTitle{Optics Letters}} \textbf{36}, 1680 (2011).

\bibitem{wittingSub4fsLaserPulse2012}
T.~Witting, F.~Frank, W.~A. Okell, C.~A. Arrell, J.~P. Marangos, and J.~W.~G.
  Tisch, \enquote{Sub-4-fs laser pulse characterization by spatially resolved
  spectral shearing interferometry and attosecond streaking,}
  {\protect\JournalTitle{Journal of Physics B: Atomic, Molecular and Optical
  Physics}} \textbf{45}, 074014 (2012).

\bibitem{weitenberg17}
J.~Weitenberg, A.~Vernaleken, J.~Schulte, A.~Ozawa, T.~Sartorius, V.~Pervak,
  H.-D. Hoffmann, T.~Udem, P.~Russb{\"u}ldt, and T.~W. H{\"a}nsch,
  \enquote{Multi-pass-cell-based nonlinear pulse compression to 115\,fs at
  7.5\,{$\mu$J} pulse energy and 300\,{W} average power,}
  {\protect\JournalTitle{Optics Express}} \textbf{25}, 20502--20510 (2017).

\bibitem{lavenu18}
L.~Lavenu, M.~Natile, F.~Guichard, Y.~Zaouter, X.~Delen, M.~Hanna, E.~Mottay,
  and P.~Georges, \enquote{Nonlinear pulse compression based on a gas-filled
  multipass cell,} {\protect\JournalTitle{Optics Letters}} \textbf{43},
  2252--2255 (2018).

\bibitem{lu18}
C.-H. Lu, T.~Witting, A.~Husakou, M.~J.~J. Vrakking, A.~H. Kung, and F.~J.
  Furch, \enquote{Sub-4 fs laser pulses at high average power and high
  repetition rate from an all-solid-state setup,} {\protect\JournalTitle{Optics
  Express}} \textbf{26}, 8941--8956 (2018).

\bibitem{daher20}
N.~Daher, F.~Guichard, S.~W. Jolly, X.~D\'{e}len, F.~Qu\'{e}r\'{e}, M.~Hanna,
  and P.~Georges, \enquote{Multipass cells: 1d numerical model and
  investigation of spatio-spectral couplings at high nonlinearity,}
  {\protect\JournalTitle{J. Opt. Soc. Am. B}} \textbf{37}, 993--999 (2020).

\bibitem{Yoon:21}
J.~W. Yoon, Y.~G. Kim, I.~W. Choi, J.~H. Sung, H.~W. Lee, S.~K. Lee, and C.~H.
  Nam, \enquote{Realization of laser intensity over $10^{23}$ w/$cm^2$,}
  {\protect\JournalTitle{Optica}} \textbf{8}, 630--635 (2021).

\bibitem{Kim:21}
Y.~G. Kim, J.~I. Kim, J.~W. Yoon, J.~H. Sung, S.~K. Lee, and C.~H. Nam,
  \enquote{Single-shot spatiotemporal characterization of a multi-pw laser
  using a multispectral wavefront sensing method,}
  {\protect\JournalTitle{Optics Express}} \textbf{29}, 19506--19514 (2021).

\bibitem{grace21}
E.~Grace, T.~Ma, Z.~Guang, R.~Jafari, J.~Park, J.~Clark, G.~Kemp, J.~Moody,
  M.~Rhodes, Y.~Ping, R.~Shepherd, B.~Stuart, and R.~Trebino,
  \enquote{Single-shot complete spatiotemporal measurement of terawatt laser
  pulses,} {\protect\JournalTitle{Journal of Optics}} \textbf{23}, 075505
  (2021).

\bibitem{gallet14}
V.~Gallet, \enquote{{Dispositifs exp\'erimentaux pour la caract\'erisation
  spatio-temporelle de chaines laser femtosecondes haute-puissance},} Ph.D.
  thesis, Université Paris XI (2014).
  \url{https://tel.archives-ouvertes.fr/tel-01084002/document}.

\bibitem{parienteCaracterisationSpatiotemporelleImpulsions2017}
G.~Pariente, \enquote{{Caractérisation spatio-temporelle d'impulsions laser de
  haute puissance},} Ph.D. thesis, Université Paris-Saclay (2017).
  \url{https://hal.archives-ouvertes.fr/tel-01487697v2/document}.

\bibitem{Tache:87}
J.-P. Tach\'{e}, \enquote{Ray matrices for tilted interfaces in laser
  resonators,} {\protect\JournalTitle{Appl. Opt.}} \textbf{26}, 427--429
  (1987).

\bibitem{Eggleston:88}
J.~M. Eggleston, L.~G. Deshazer, and K.~W. Kangas, \enquote{Characteristics and
  kinetics of laser-pumped {Ti:}sapphire oscillators,}
  {\protect\JournalTitle{IEEE Journal of Quantum Electronics}} \textbf{24},
  1009--1015 (1988).

\bibitem{Ohta:88}
K.~Ohta and H.~Ishida, \enquote{Comparison among several numerical integration
  methods for {Kramers-Kronig} transformation,} {\protect\JournalTitle{Appl.
  Spectrosc.}} \textbf{42}, 952--957 (1988).

\bibitem{Malitson:62}
I.~H. Malitson, \enquote{Refraction and dispersion of synthetic sapphire,}
  {\protect\JournalTitle{J. Opt. Soc. Am.}} \textbf{52}, 1377--1379 (1962).

\bibitem{McLeod:14}
R.~R. McLeod and K.~H. Wagner, \enquote{Vector {Fourier} optics of anisotropic
  materials,} {\protect\JournalTitle{Adv. Opt. Photon.}} \textbf{6}, 368--412
  (2014).

\bibitem{Morice:1999}
O.~Morice, X.~Ribeyre, and V.~Rivoire, \enquote{{Broad-band computations using
  the Mir\`o{} software},} in \emph{Third International Conference on Solid
  State Lasers for Application to Inertial Confinement Fusion,}  vol. 3492
  W.~H. Lowdermilk, ed., International Society for Optics and Photonics (SPIE,
  1999), pp. 832 -- 838.

\bibitem{Paschotta:17}
R.~Paschotta, \enquote{Modeling of ultrashort pulse amplification with gain
  saturation,} {\protect\JournalTitle{Opt. Express}} \textbf{25}, 19112--19116
  (2017).

\end{thebibliography}

\end{document}